\documentclass[showpacs,amsmath,amssymb,aps,twocolumn,superscriptaddress,prb]{revtex4-2} 

\usepackage[colorlinks=true,citecolor=blue,linkcolor=magenta]{hyperref}
\usepackage[usenames]{color}
\usepackage{relsize}
\usepackage[english]{babel}
\usepackage{enumerate}

\usepackage{graphicx}
\graphicspath{{./},{figs_main/},{figs_SI/},{figs_App/},}

\newcommand{\SI}{Supplemental Material}

\newcommand {\grsim} {\ {\raise-.5ex\hbox{$\buildrel>\over\sim$}}\ }
\newcommand {\lessim} {\ {\raise-.5ex\hbox{$\buildrel<\over\sim$}}\ }

\usepackage{grffile} 

\usepackage{comment}

\usepackage{float}
\usepackage{dcolumn}
\usepackage{bm}

\newcommand{\eqn}[1]{\begin{equation} #1 \end{equation}}
\newcommand{\eqa}[1]{\begin{align} #1 \end{align}}

\usepackage{color}
\definecolor{ForestGreen}{RGB}{34, 139, 34}

\newcommand{\nn}{\nonumber}

\newcommand{\mH}{\mathcal{H}}

\newcommand{\mV}{\mathcal{V}}
\newcommand{\avg}[1]{\left\langle #1 \right\rangle}
\newcommand{\pd}{\partial}
\newcommand{\bS}{\boldsymbol{S}}
\newcommand{\bM}{\boldsymbol{M}}

\newcommand{\ba}{\boldsymbol{a}}

\newcommand{\bEta}{\boldsymbol{\eta}}
\newcommand{\bsigma}{\boldsymbol{\sigma}}

\newcommand{\mO}{\mathcal{O}}

\newcommand{\mA}{\mathcal{A}}

\begin{document}
	
\def \titletext {Prethermalization in periodically-driven nonreciprocal many-body spin systems}

\def \abstracttext {
	We analyze a new class of time-periodic nonreciprocal dynamics in interacting chaotic classical spin systems, whose equations of motion are conservative (phase-space-volume-preserving) yet possess no symplectic structure. As a result, the dynamics of the system cannot be derived from any time-dependent Hamiltonian.
	In the high-frequency limit, we find that the magnetization dynamics features a long-lived metastable plateau, whose duration is controlled by the fourth power of the drive frequency. However, due to the lack of an effective Hamiltonian, the prethermal state the system evolves into cannot be understood within the framework of the canonical ensemble. We propose a Hamiltonian extension of the system using auxiliary degrees of freedom, in which the original spins constitute an open yet nondissipative subsystem. This allows us to perturbatively derive effective equations of motion that manifestly display symplecticity breaking at leading order in the inverse frequency. 
	We thus extend the notion of prethermal dynamics, observed in the high-frequency limit of periodically-driven systems, to nonreciprocal systems. 
}

\author{Adam J. McRoberts}
\email{amcr@pks.mpg.de}
\affiliation{Max Planck Institute for the Physics of Complex Systems, N\"{o}thnitzer Str.~38, 01187 Dresden, Germany}

\author{Hongzheng Zhao}
\affiliation{Max Planck Institute for the Physics of Complex Systems, N\"{o}thnitzer Str.~38, 01187 Dresden, Germany}

\author{Roderich Moessner}
\affiliation{Max Planck Institute for the Physics of Complex Systems, N\"{o}thnitzer Str.~38, 01187 Dresden, Germany}

\author{Marin Bukov}
\email{mgbukov@phys.uni-sofia.bg}
\affiliation{Max Planck Institute for the Physics of Complex Systems, N\"{o}thnitzer Str.~38, 01187 Dresden, Germany}
\affiliation{Department of Physics, St.~Kliment Ohridski University of Sofia, 5 James Bourchier Blvd, 1164 Sofia, Bulgaria}
	
\title{\titletext}

\begin{abstract} 
	\abstracttext
\end{abstract}

\maketitle



\section{Introduction}

Systems exhibiting nonreciprocal interactions evade Newton's third law, and are intrinsically out-of-equilibrium due to the absence of energy conservation.
As an effective description of physical systems, nonreciprocity finds a plethora of applications ~\cite{bowick2022symmetry}, underlying flocking phenomena in active matter~\cite{barberis2016large,dadhichi2020nonmutual}, 
interactions between microparticles in an anisotropic plasma~\cite{lisin2020experimental},
and the formation of active chiral matter in starfish embryos.
Outwith biophysics, such interactions arise naturally in systems of colloidal particles interacting through nonreciprocal electrostatic torques, tuneable by changing either the salt concentration or the external electric field~\cite{yan2016reconfiguring,zhang2021active}; moreover, nonreciprocal interactions have been shown to exhibit out-of-equilibrium phase transitions~\cite{fruchart2021non} and non-Hermitian topology~\cite{ghatak2020observation}, and have recently been emulated between robots in a programmable way~\cite{brandenbourger2021limit}.

\begin{figure}[t!]
	\centering
	\includegraphics[width=\columnwidth]{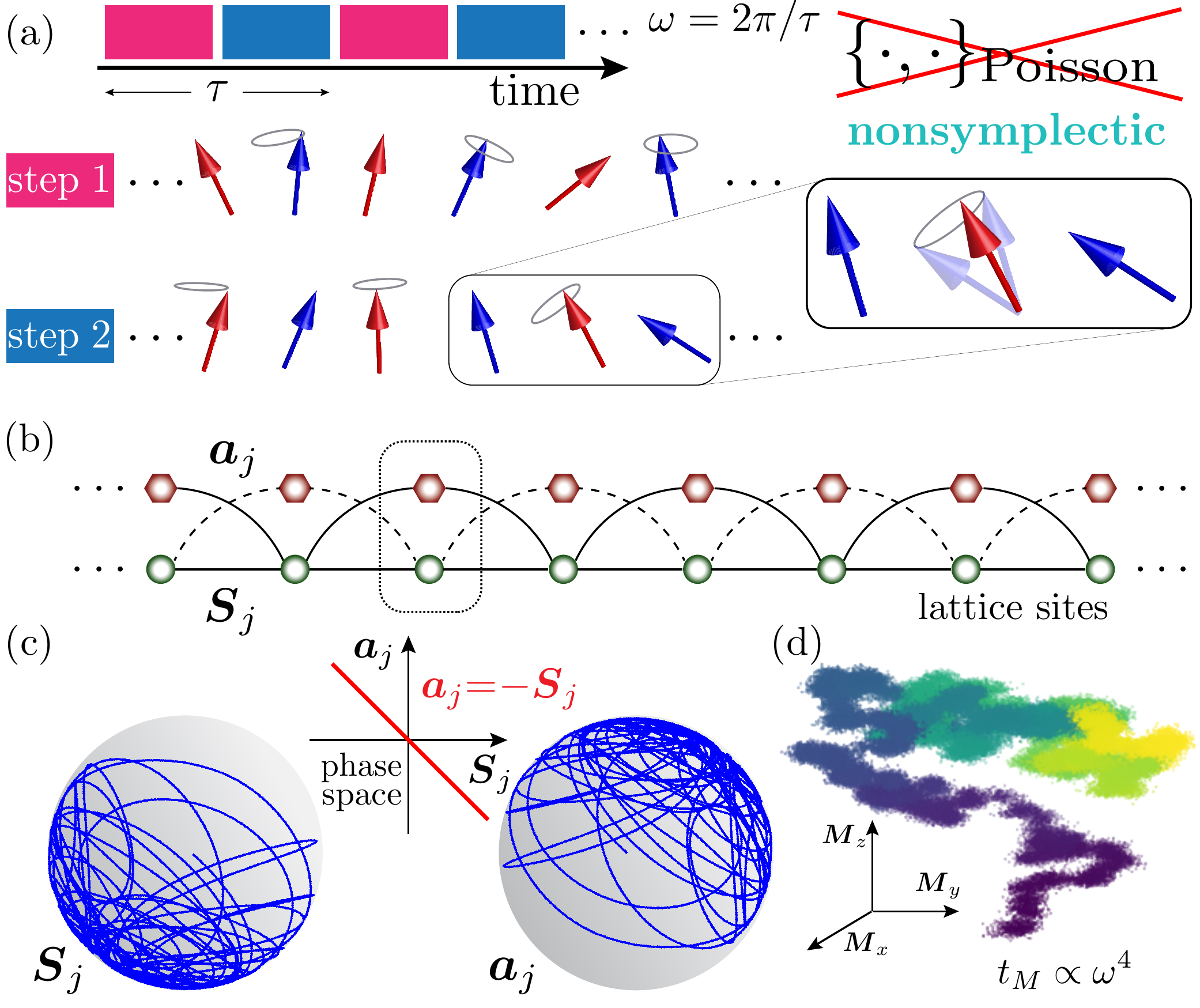}
	\caption{
		(a) An interacting nonintegrable spin chain is subject to time-periodic two-step dynamics that breaks the reciprocity of the interactions and the symplectic structure of phase space. In the first half-cycle spins on one sublattice are held fixed, while the other spins precess in the exchange field of their neighbors, and vice-versa during the second half-cycle. 
		(b) The spin chain can be embedded in a larger Hamiltonian system comprised of two interacting dynamical spin degrees of freedom $\bS_j$ and $\ba_j$, which restores symplecticity [see text]. 
		(c) The initial condition $\ba_j{=-}\bS_j$ confines the phase space dynamics of the $(\bS,\ba)$-system to a subspace for all time; the $\bS$-subsystem is exactly described by the nonsymplectic drive from (a). 
		(d) The nonreciprocal drive breaks magnetization conservation: at large drive frequencies $\omega$, the magnetization of an initial ensemble (shown by collection of points) exhibits a prethermal plateau whose lifetime scales as $t_M\sim\omega^4$, before it relaxes in a diffusion-like process (colorcode shows arrow of time from purple to yellow). 
	}
	\label{fig:Geometry}
\end{figure}

In this article, we investigate the consequences of nonreciprocity for the out-of-equilibrium dynamics of spin systems subject to a time-periodic drive.
Floquet-driven, \textit{Hamiltonian}, closed, many-body systems generically feature long-lived metastable states in the high-frequency regime, when drive frequencies are much larger 
than the local bandwidth~\cite{abanin2015exponentially,mori2016rigorous}. For sufficiently short-ranged interactions, such systems feature exponentially long-lived prethermal plateaus, where energy absorption is severely constrained and slowed down, as higher-order interaction processes are required~\cite{lazarides2014equilibrium,dumitrescu2018logarithmically,machado2019exponentially,else2020long,zhao2021random,fleckensteon2021prethermal,kyprianidis2021observation}. Experimentally, Floquet prethermalization has been instrumental for the realization of novel engineered properties~\cite{goldman2014periodically,bukov2015universal,moessner2017equilibration,eckardt2017colloquium,oka2019floquet}, such as artificial gauge fields for neutral particles~\cite{struck2013engineering,aidelsburger2013realization,schweizer2019floquet}; discrete time crystalline~\cite{khemani2016phase,else2016floquet,yao2017discrete,pizzi2021classicalprethermal,beatrez2022observation} or topologically-ordered~\cite{potirniche2017floquet,wintersperger2020realization,decker2020floquet} phases of matter without equilibrium counterparts; as a stabilization mechanism to create long-lived coherent dynamics~\cite{singh2019quantifying,rubio2020floquet,haldar2021dynamics,peng2021floquet,beatrez2021Floquet,sahin2022continuously};
and in providing a long time-window to realize Trotterized dynamics on digital quantum processors~\cite{heyl2019quantum}. For all these reasons, Floquet drives offer a highly versatile toolbox; however, the extent to which \textit{nonreciprocal} systems are amenable to Floquet engineering remains unclear~\cite{koutserimpas2018nonreciprocal,li2018floquet}.

In this article, we ask whether closed, \textit{nonreciprocal} many-body systems, subject to a periodic, conservative (i.e., phase-space-volume-preserving) drive, can exhibit long-lived prethermalization -- in other words, can nonreciprocal systems offer a suitable framework to implement ideas from Floquet engineering?

We give an affirmative answer by investigating the magnetization relaxation of a classical many-body spin chain~\cite{howell2019asymptotic,mori2018floquet,notarnicola2018from,rajak2019characterizations,jin2022prethermal,kundu2021dynamics,ye2021floquet,sadia2022prethermalization} exposed to such a drive [Fig.~\ref{fig:Geometry}a]. Unlike their 
Hamiltonian Floquet counterparts~\cite{higashikawa2018floquet,mori2018floquet}, nonreciprocal periodically-driven equations of motion (EOM) cannot be derived from any time-dependent Hamiltonian, and, therefore, possess no well-defined quasienergy. As a consequence, they cannot, \textit{a priori}, exhibit energy prethermalization and are not described by an effective Hamiltonian, even in the high-frequency regime. 

Nevertheless, we show that nonreciprocal time-periodic dynamics can exhibit quasi-conserved quantities which relax through a parametrically controlled long-lived prethermal plateau, the duration of which scales as the \textit{fourth power} of the drive frequency. We 
derive an effective stroboscopic EOM in the high-frequency regime, by considering the spin chain as a subsystem of a larger Hamiltonian system 
[Fig.~\ref{fig:Geometry}b]; the leading-order inverse-frequency correction is sufficient to capture the magnetization relaxation process [Fig.~\ref{fig:Geometry}d].
The nonreciprocal periodic drive we investigate is applicable to various classical spin models
, irrespective of their dimensionality, support of interactions, and lattice geometry, and thus defines a distinct class of prethermal states.


\section{Model}

We consider a bipartite lattice of interacting classical spins $\bS_j$, governed by the time-periodic EOM
\begin{eqnarray}
	\label{eq:EOM}
	\begin{array}{lr}
		\begin{cases}
			\dot\bS_j^{\mu} = \epsilon^{\mu\nu\lambda} \left(\sum_i J_{ij}^{\nu}\bS_{i}^{\nu} \right)\bS_j^{\lambda},\quad & j \in \mathcal{A}\\
			\dot\bS_j^{\mu} =  0,\quad & j \in \mathcal{B}
		\end{cases};\quad
		&\text{for}\; t\in\left[0,\frac{\tau}{2}\right), \\ \\
		\begin{cases}
			\dot\bS_j^{\mu} = 0,\quad & j \in \mathcal{A}\\
			\dot\bS_j^{\mu} = \epsilon^{\mu\nu\lambda} \left(\sum_i J_{ij}^{\nu}\bS_{i}^{\nu} \right)\bS_j^{\lambda},\quad & j \in \mathcal{B}
		\end{cases};\quad
		&\text{for}\; t\in\left[\frac{\tau}{2},\tau\right), 
	\end{array}
\end{eqnarray}
where $J_{ij}^{\nu} = J_{ji}^{\nu}$ denotes the interaction strength \footnote{Despite the symmetry of the interaction coupling, $J_{ij}^{\nu} = J_{ji}^{\nu}$, the interactions between the spins are nonreciprocal during each half-step by construction.}, $i, j$ label the lattice sites, and $\mathcal{A}, \mathcal{B}$ are the two sublattices; in all simulations, we use periodic boundary conditions.
 
During the first (second) half-cycle, the spins on the $\mathcal{B}$ ($\mathcal{A}$) sublattice are kept frozen, and produce an effective constant field in which their neighboring spins precess [Fig.~\ref{fig:Geometry}a]; the roles of the two sublattices are then flipped, and the protocol repeats.
Since the rotation axis depends on the neighboring spins -- the directions of which keep changing -- this protocol gives rise to chaotic nonlinear dynamics over many drive periods $\tau$; the frequency of switching is $\omega=2\pi/\tau$. 

We define the infinite-frequency limit by fixing a physical time in units of $J^{-1}$, and solving the EOM up to that time as $\tau \rightarrow 0$.
In this limit, averaging over a period reduces Eq.~\eqref{eq:EOM} to the familiar Bloch equations, $\dot\bS_j^{\mu} = \{\bS_j^{\mu},H_\mathrm{\infty}\}$, generated by the Hamiltonian,
\eqn{
	H_\mathrm{\infty} = \frac{1}{2} \sum_{i,j} J_{ij}^{\mu}\bS_i^{\mu} \bS_j^{\mu},
	\label{eq:H_Heisenberg}
}
where $\{\cdot,\cdot\}$ denotes the Poisson bracket, with $\{\bS^\alpha_i,\bS^\beta_j\}=\delta_{ij}\varepsilon^{\alpha\beta\gamma}\bS_j^\gamma$. In all cases, we assume an O(2) isotropy, $J_{ij}^x = J_{ij}^y$. The infinite-frequency dynamics is thus Hamiltonian, and conserves both the magnetization $\bM_z = \sum_j \bS_j^z$ and the infinite-frequency energy $H_\mathrm{\infty}$. At finite frequency, $H_\mathrm{\infty}$ remains conserved for all time, but the magnetization is no longer conserved. 

A key feature of Eq.~\eqref{eq:EOM} is that, for finite drive frequencies, it cannot be derived from any Hamiltonian, time-dependent or static, since nonreciprocity breaks the symplectic structure~[App.~\ref{app:non_sympl}].
Nevertheless, it is curious to note that Eq.~\eqref{eq:EOM} still preserves the phase space volume~\footnote{The phase space volume is preserved independently during each half-cycle, and hence also by the entire dynamics.}, and hence the dynamics remains incompressible at all frequencies. 

The lack of a Hamiltonian implies that the dynamics of Eq.~(\ref{eq:EOM}) is not directly amenable to Floquet theory [but see Ref.~\cite{higashikawa2018floquet}]. This raises two questions: 
(i) what are the similarities and differences between nonreciprocal and Hamiltonian Floquet systems, and
(ii) how can we effectively describe their thermalizing dynamics?


\section{Long-Lived Metastable Plateau}

We address these questions numerically
before presenting a theoretical description. 
We initialize the chain in a canonical ensemble at temperature $\beta^{-1}$, magnetized along the $z$-direction; more specifically, this ensemble is thermal w.r.t.~the Hamiltonian $\tilde{H} = 2H_\mathrm{\infty} + h\sum_j \bS^z_j$, at $h/|J| = 0.7$.
We then evolve each state in the ensemble up to a sufficiently long time, and measure the expectation value (ensemble-average) of the magnetization.

Since $H_\mathrm{\infty}$ is conserved for each state in the ensemble, the system cannot heat w.r.t.~$H_\mathrm{\infty}$ -- nevertheless, it can still absorb energy w.r.t.~the Hamiltonian $\tilde{H}$ that generated the initial ensemble; this gives rise to magnetization relaxation, and we therefore colloquially refer to the dynamics as ``thermalizing". 

\begin{figure}[t!]
	\centering
           \includegraphics[width=\columnwidth]{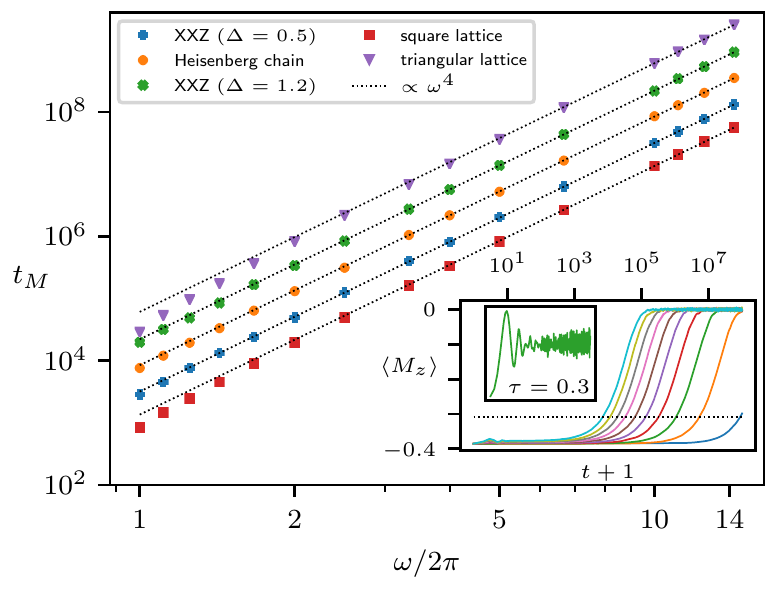}
	\caption{
		Magnetization relaxation in the nonreciprocal drive, Eq.~\eqref{eq:EOM}, occurs via a prethermal plateau whose lifetime is controlled by the drive frequency $\omega$. The models shown are the XXZ spin chain ($J^x {=} J^y {=} 1$, $J^z {=} \Delta$) and the Heisenberg model ($J^x {=} J^y {=} J^z {=} 1$) on the chain, square lattice, and triangular lattice.
  The inset shows the relaxation curves for the Heisenberg chain for drive periods $\tau = 0.1$ (blue, rightmost) to $1.0$ (cyan, leftmost), in steps of $0.1$  -- their intersection with the horizontal line defines $t_M$. The main plot shows that the relaxation timescale $t_M {\sim} \omega^4$ scales with the fourth power of the drive frequency $\omega$ (dotted lines), regardless of the particular spin model. The data have been shifted vertically for clarity. All ensemble averages are taken over $2000$ initial states, at temperature $\beta = 1$ and initial applied field $h_z {=} 0.7$. The system size for the chains is $L {=} 128$, and the linear sizes for the 2D lattices are $L {=} 16$ (square) and $L {=} 15$ (triangular). The nonreciprocal drive for the triangular lattice does not follow Eq.~\eqref{eq:EOM} (it is not bipartite), but follows an analogous three-step protocol [App.~\ref{app:other_systems}].
	}
	\label{fig:sdcq_M}
\end{figure}

To quantify the rate at which this happens as a function of $\omega$, we study the magnetization relaxation for several spin models: the XXZ spin chain ($J^x {=} J^y {=} 1$, $J^z {=} \Delta$), for both easy-axis and easy-plane anisotropy, and the nearest-neighbor isotropic Heisenberg model ($J^x {=} J^y {=} J^z {=} 1$) on the chain, square lattice, and triangular lattice. 
Figure~\ref{fig:sdcq_M} [inset] shows the time evolution of $\langle \bM_z\rangle$ for different values of the drive frequency. 
Despite the lack of symplecticity in Eq.~\eqref{eq:EOM}, the overall behavior appears similar to Hamiltonian Floquet drives:
the system first prethermalizes to a frequency-dependent plateau above the infinite-frequency magnetization value, which lasts until a time $t_M$, parametrically controlled by $\omega$, when the ensemble starts approaching the $\langle \bM_z\rangle = 0$ state. 
Since no effective Hamiltonian exists that describes the prethermal plateau, there exists no canonical ensemble description of the state and no effective temperature can be assigned to it. Therefore, nonreciprocal prethermal states should be understood within the microcanonical ensemble as maximum-entropy states on the accessible phase-space manifold.
Curiously, however, we observe a power-law scaling $t_M\propto\omega^{\alpha}$ with $\alpha=4$, \textit{independently of the particular spin model}
. This is in stark contrast to both the exponential scaling 
in locally-interacting Hamiltonian Floquet systems~\cite{mori2018floquet}, and the $\alpha=2$ Fermi's Golden rule regime characteristic of long-range systems~\cite{machado2019exponentially}. We emphasize that the proofs of rigorous upper bounds on energy absorption explicitly use the Hamiltonian formalism~\cite{abanin2015exponentially,mori2016rigorous,ho2018bounds}, and hence do not apply to nonreciprocal dynamics. 

To contrast the nonreciprocal drive from its reciprocal counterpart, let us specialize to the isotropic Heisenberg chain, ${H_{\infty} = \frac{J}{2}\sum_j \bS_j\cdot\bS_{j+1}}$, and consider the Hamiltonian Floquet drive $H(t) {=}\sum_{j} J_j(t)\bS_j{\cdot}\bS_{j+1}$ with $J_j(t){=}{J}/{2}\left[1{+}(-1)^j\mathrm{sgn}\left(\sin\omega t\right)\right]$. Both drives share the same infinite-frequency Hamiltonian $H_\mathrm{\infty}$ -- at finite frequency, however, the Hamiltonian structure implies the heating timescale is exponentially suppressed [App.~\ref{app:bond_drive}], in accordance with the theorem of Ref.~\cite{mori2018floquet}.


\section{Effective Description}

The numerical observation of prethermalization calls for a theoretical description;
this is complicated, however, by the fact that nonreciprocity precludes the description of the spin dynamics using an effective Floquet Hamiltonian.
We sidestep this issue by explicitly constructing a larger, Hamiltonian (i.e., reciprocal) system, of which the original spins constitute an open but nondissipative subsystem.

Each half-cycle of the dynamics governed by Eq.~\eqref{eq:EOM} can be realized in the presence of local external magnetic fields that cancel the precession of the static spins. For this to happen during both half-cycles, the external fields have to change both their direction and lattice support in time; we thus need to promote them to dynamical degrees of freedom. 
We introduce an auxiliary spin $\ba_j$ on every lattice site, which obeys $\{\ba_i^\alpha, \ba_j^\beta\}=\varepsilon^{\alpha\beta\gamma}\ba^\gamma_j\delta_{ij}$ and $\{\ba_i^\alpha, \bS_j^\beta\}=0$, cf.~Fig.~\ref{fig:Geometry}b. Each $\ba_j$ couples periodically to the neighbors of $\bS_j$, giving rise to the Hamiltonian
\begin{equation}
	\label{eq:H-tot}
	\mathcal{H}(t) = \sum_{i,j} J_{ij}^{\mu}\bS_i^{\mu}\bS_j^{\mu} \!+\! \left[\frac{1}{2}\!+\!g(t)\mathrm{sgn}(j)\right] J_{ij}^{\mu}\bS_i^{\mu}\ba_j^{\mu},
\end{equation}
where $g(t) = \;\mathrm{sgn}(\sin\omega t)/2$ is a $\tau$-periodic step-drive, and $\mathrm{sgn}(j)$ takes different signs on the two sublattices. 

In general, the chaotic dynamics generated by the total Hamiltonian~\eqref{eq:H-tot} differs from Eq.~\eqref{eq:EOM}. One may show, however, that an initial condition of the form ${\ba_j(0) = -\bS_j(0)}$ is preserved, i.e., ${\ba_j(t) = -\bS_j(t)}$ ~[App.~\ref{app:eff_EOM}], and, under these conditions, we recover exactly the EOM for the original spin chain from Eq.~\eqref{eq:EOM}.

The Hamiltonian~\eqref{eq:H-tot} sheds new light on our problem, as it allows us to think of the $\bS$-spins as an open system. Note, however, that the dynamics described by Eq.~\eqref{eq:EOM} is conservative, since the Poincar\'e recurrence theorem is satisfied~\footnote{One can convince oneself that both conditions for Poincar\'e's recurrence theorem are met for any fixed number of spins: the phase space flow is incompressible, cf.~App.~\ref{app:non_sympl} and all orbits are bounded since the spin phase space is compact.}. 

Adopting this view, we find that, for $\ba_j(t) \!=\! -\bS_j(t)$, the total energy of the system vanishes identically, ${\mathcal{H}(t) \equiv 0}$. Since the energy of the $\bS$-spin subsystem is independently conserved, it follows that the energy absorbed from the periodic drive in the total system remains trapped in the interaction term between the two systems.  
Moreover, the magnetization of the total system also vanishes identically, $\sum_j \bS_j\!+\!\ba_j\!\equiv\!0$. Thus, we can interpret the slow magnetization relaxation in the high-frequency limit as scrambling dynamics within the $\bM{=}0$ shell of the full system. 

Let us emphasize that the specific choice of initial condition for the total system should not be viewed as fine-tuning. Indeed, we are interested in the dynamics of the $\bS$-subsystem, for which the initial condition is arbitrary; the $\ba$-subsystem helps us to make analytic progress in a familiar and structured way. 
Due to the nonlinearity of the EOM generated by Eq.~\eqref{eq:H-tot}, we expect that small deviations from this initial condition lead to unstable dynamics that leaves the ${\ba=-\bS}$ manifold and features fundamentally different properties.

\section{Floquet-Magnus Expansion}

The Hamiltonian (\ref{eq:H-tot}) enables the application of Floquet theory. Since prethermalization requires high frequencies, let us consider the inverse-frequency expansions (IFE). 
For the purpose of deriving an effective EOM, it suffices to focus on the stroboscopic dynamics which governs the motion of the slow degrees of freedom.  Out of the different variants, we choose the Floquet-Magnus expansion since it does not require kick operators that modify the initial conditions~\footnote{Classically, kick operators give rise to canonical transformations that are more difficult to apply, as compared to unitary change of basis in quantum mechanics. The latter is the case for the van-Vleck expansion~\cite{bukov2015universal}.}.

From this point, we will, for concreteness, focus on the isotropic Heisenberg chain. A straightforward calculation of the Floquet Hamiltonian yields [App.~\ref{app:eff_EOM}]
\begin{eqnarray}
	\label{eq:H_F}
	\mH_F &=& \mH_F^{(0)} + \mH_F^{(1)} +\mathcal{O}(\omega^{-2}),\quad \mH_F^{(n)}\propto\omega^{-n},\nonumber\\
	\mH_F^{(0)} &=& J \sum_{j}\bS_{j} \cdot \bS_{j+1}+ \frac{1}{2} \ba_{j} \cdot\left(\bS_{j-1}+\bS_{j+1}\right)  ,\nonumber\\
	\mH_F^{(1)} &=& -\frac{J^{2}\tau}{8} \sum_{j}(-1)^{j} \ba_{j} \cdot\bigg[\left(\bS_{j} +\bS_{j-2}\right) \times \bS_{j-1}\nonumber\\
	&&\qquad\qquad\qquad\qquad+\left(\bS_{j+2}+\bS_{j}\right) \times \bS_{j+1}\bigg].
\end{eqnarray}
The zeroth-order term is the period-averaged Hamiltonian, and includes static interactions between the original and the auxiliary spins. The first-order correction contains $\ba_j$-mediated nearest-neighbor interactions between the spins $\bS_j$, and breaks time-reversal symmetry.


\section{Effective Stroboscopic Dynamics}

Equation~\ref{eq:H_F} immediately yields the stroboscopic equations of motion, $\dot\ba_j \!=\! \{\ba_j,\mH_F^{(0+1)}\}$ and $\dot\bS_j \!=\! \{\bS_j,\mH_F^{(0+1)}\}$.
Again, it may be shown that ${\ba_j(0) \!=\! -\bS_j(0)}$ implies ${\ba_j(t) \!=\! -\bS_j(t)}$ within the effective dynamics [App.~\ref{app:eff_EOM}]. Using this to eliminate the $\ba$-spins, we obtain an effective EOM for the $\bS$-spins:
\begin{eqnarray}
	\label{eq:EOM-eff}
	\dot{\bS}_{j} &=& \frac{J}{2}\left(\bS_{j-1}+\bS_{j+1}\right)\times \bS_{j} - \frac{J^{2}\tau }{8}(-1)^{j}\cdot
	\\
	&&\Big[\left(\bS_{j}+\bS_{j-2}\right) \times \bS_{j-1}
	+\left(\bS_{j+2}+\bS_{j}\right) \times \bS_{j+1}\Big] \times \bS_{j}. \nonumber
\end{eqnarray}
As expected, the zeroth-order term corresponds to the time-averaged Heisenberg dynamics, Eq.~\eqref{eq:H_Heisenberg}. More interestingly, the first-order terms $(\bS_j\times\bS_{j\pm1})\times\bS_j$ represent non-Hamiltonian corrections; they cannot be derived from any $\bS$-subsystem Hamiltonian $H_\mathrm{eff}$ via $(\partial H_\mathrm{eff}/\partial\bS_j) \times \bS_j$ [App.~\ref{app:nonsymplecticEOM_eff}].
Symplecticity is thus already broken at leading order in $\omega^{-1}$.
On the other hand, the $\mathcal{O}(\omega^{-1})$ dynamics conserves the infinite-frequency energy $H_{\mathrm{\infty}}$. Since the exact dynamics also conserves $H_{\mathrm{\infty}}$, we conjecture that this is true at all orders in the IFE.

\begin{figure}[t!]
	\centering
           \includegraphics{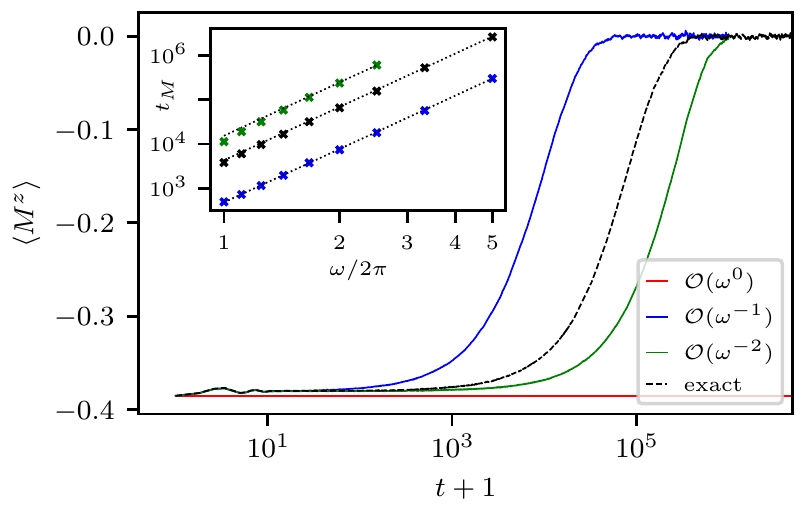}
	\caption{
		Comparison between the effective stroboscopic dynamics and the exact dynamics in the isotropic Heisenberg chain. 
		Main figure shows the magnetization dynamics within the first three orders of the inverse-frequency expansion (IFE): $M^z$ is conserved at zeroth-order; the first-order curve relaxes faster than the exact curve; whilst the second-order curve takes longer, hinting at a possible oscillatory convergence. Even at first-order, the IFE captures the short-time quenched dynamics that drives the state into the `prethermal' plateau. 
		The inset shows that the power-law scaling of the relaxation time, $t_M \sim \omega^{4}$, (dotted lines) is also captured at first-order in the IFE. 
		Simulation parameters are the same as in Fig.~\ref{fig:sdcq_M}, with $\beta=1$. The curves in the main figure correspond to $\tau = 0.8$.
	}
	\label{fig:effective_dynamics}
\end{figure}

However, the $\omega^{-1}$-correction breaks magnetization conservation. Figure~\ref{fig:effective_dynamics} shows a comparison between the exact and effective magnetization dynamics up to and including $\mathcal{O}(\omega^{-2})$. Note that magnetization relaxation, including the prethermal plateau, is already captured by the $\mO(\omega^{-1})$ terms. This contrasts with energy relaxation in 
Hamiltonian Floquet systems [App.~\ref{app:bond_drive}], where the effective dynamics does not capture heating to infinite temperature at any order~\cite{bukov2016heating}, and the Floquet-Magnus expansion diverges~\cite{mori2016rigorous}. 
Curiously, we find that the $\mO(\omega^{-1})$ dynamics relaxes magnetization faster than the exact dynamics. This is peculiar, since taking into account all higher-order corrections (i.e., considering the exact dynamics) adds more long-range and multi-body terms to the effective EOM, which one would expect to lead to faster relaxation. 
At the same time, our analysis reveals that the scaling of the relaxation time is ${t_M\sim\omega^{4}}$ in both the first- and second-order effective EOM [Fig.~\ref{fig:effective_dynamics} inset]; the differences in relaxation times are, therefore, caused by a nonuniversal truncation-order-dependent prefactor $c^{(n)}$:
$t^{(n)}_M{=}c^{(n)}\omega^4$.
We observe that the second-order dynamics leads to magnetization relaxation that is slower than the exact dynamics, hinting at the possibility of an oscillatory convergence to the exact magnetization curve as we include higher-order terms of the IFE~\footnote{Fewer data points are available at second-order, since the slower relaxation makes determining $t_M$ infeasible for $\omega/2\pi \gtrsim 3$ over the accessible timescales $t \sim 10^6$.}.

The origin of the particular exponent, $\alpha = 4$, governing the prethermal lifetime remains unclear; however, it cannot be explained perturbatively using the IFE: Instead of the Floquet-Magnus IFE, one might be tempted to consider the van Vleck IFE~\cite{bukov2015universal} -- since the latter is manifestly independent of the initial time; for a linearly polarized, real-valued Hamiltonian $\mathcal{H}(t)$, a straightforward calculation shows that the $\mO(\omega^{-1})$ terms vanish, and hence ${\mathcal{H}_\mathrm{eff}\sim\mathcal{H}_\mathrm{eff}^{(0)}+\mathcal{H}_\mathrm{eff}^{(2)}}$. In the high-frequency regime, ${\mathcal{H}_\mathrm{eff}^{(2)}\in\mathcal{O}(\omega^{-2})}$ would then play the role of a weak $\bS$-magnetization breaking perturbation; a Fermi's golden rule-type argument would then imply that the magnetization relaxation rate cannot be faster than $\Gamma_M{\sim}(\omega^{-2})^2$, the inverse of which defines the prethermal lifetime scaling as seen in Fig.~\ref{fig:sdcq_M}. 

However, we demonstrate explicitly in App.~\ref{app:other_systems} that such a na\"ive perturbative argument based on the IFE is incorrect: the nonreciprocal dynamics of any time-reversal-breaking periodic drive has a non-vanishing first-order correction ${\mathcal{H}_\mathrm{eff}^{(1)}\in\mathcal{O}(\omega^{-1})}$, which, by the above argument, would predict a prethermal lifetime $t_M \sim \omega^2$. 
The three-step drive for the triangular lattice -- where, analogously to Eq.~\eqref{eq:EOM}, we evolve the spins on the three sublattices in an $\mathcal{A}\mathcal{B}\mathcal{C}\mathcal{A}\mathcal{B}\mathcal{C}$ pattern -- breaks time-reversal symmetry because there is no time $t_0$ about which this drive is even; time-reversal will always flip this to a $\mathcal{CBACBA}$ pattern (by contrast, the bipartite drive \eqref{eq:EOM} \textit{is} time-reversal symmetric about $t_0 = \tau/4$). Consequently, its effective Hamiltonian has a non-vanishing first-order correction [App.~\ref{app:other_systems}], but, as shown in Fig.~\ref{fig:sdcq_M}, the triangular lattice unambiguously evinces the same $\omega^4$-scaling as the bipartite models.

Finally, we mention that a Hamiltonian description is not required to derive the effective EOM [App.~\ref{app:alt_deriv_EOM}]. We have independently derived Eq.~\eqref{eq:EOM-eff} by using two-times perturbation theory and the phase-space density approach. We believe that an alternative Lagrangian description exists as well~\cite{elbracht2020topological}.


\section{Discussion \& Outlook}

In summary, we have identified a novel class of prethermalizing dynamics in classical periodically-driven spin systems, characterized by conservative, nonreciprocal chaotic dynamics; the systems we study thus differ from conventional periodically-driven \textit{dissipative} systems.
The long-time behavior of the magnetization dynamics features a prethermal plateau, whose lifetime scales as the fourth power of the drive frequency. 
By considering the spins to be part of a larger system, we have derived an approximate description for the effective dynamics using the IFE, which captures the magnetization relaxation. 


Our extended model $\mathcal{H}(t)$ can be viewed as an example of nonergodic scarred dynamics~\cite{bernien2017probing,turner2018weak,serbyn2021quantum,su2022observation} in classical many-body systems: the attainable phase space of the total $(\bS,\ba)$-system is constrained via the initial condition for all time [Fig.~\ref{fig:Geometry}c]. It will be intriguing to explore the information spreading in such constrained systems~\cite{sala2020ergodicity,zhao2021orthogonal,hahn2021information,pizzi2022bridging,deger2022arresting}. 
Other interesting directions that go beyond the strict periodicity of the drive include generalizations to random multipolar driving and quasiperiodic extensions~\cite{zhao2021random,cai20221,bhattacharjee2021quasi,long2022many}.

Furthermore, our analysis is in practice directly related to classical ODE solvers designed to conserve integrals of motion exactly. 
Prethermalization establishes the parametric stability of symplectic integrators~\cite{steinigeweg2006symplectic,frank1997geometric}, which conserve certain integrals of motion exactly~\cite{yoshida1993recent}; however, energy conservation is usually lost~\cite{zhong1988lie}.
While nonsymplectic, energy-conserving integration schemes can be implemented instead~\cite{faou2004energy}, the analysis of their stability is confounded by the absence of conjugate variables and Poisson brackets -- conventionally required to apply the high-frequency expansion in the analysis of higher-order heating processes.
Similar to recent work on Floquet Trotterization in quantum systems~\cite{heyl2019quantum}, the analysis of prethermal plateaus can improve the techniques for simulating equations of motion, allowing us to probe the hydrodynamic regimes of these systems with simulations of larger systems and longer times. 
Thus, our work establishes a direct relation between the thermalizing dynamics of nonreciprocal systems and the accuracy of nonsymplectic integrators.


Finally, this work demonstrates that Floquet engineering can be used beyond Hamiltonian systems. 
Particularly interesting in this context is the possibility to suppress leading-order non-Hamiltonian corrections using model parameters, and engineer quasi-conservation laws in nonreciprocal dynamical systems. This sheds new light on the applicability of the Floquet toolbox to, e.g., biophysics, where systems without a Hamiltonian description are abundant. 

\textit{Acknowledgments.---}We are grateful to R.~Alert, V.~Bulchandani, A.~Chandran, P.~Claeys, 
M.~Heyl, J.~Knolle, F.~Mintert, V.~Oganesyan,
A. Pizzi, and T.~Prosen for inspiring discussions. 
M.B.~was supported by the Marie Sk\l{}odowska-Curie grant agreement No 890711. This work was in part supported by the Deutsche Forschungsgemeinschaft  under grants SFB 1143 (project-id 247310070) and the cluster of excellence ct.qmat (EXC 2147, project-id 390858490).


\bibliography{References}


%
%

\cleardoublepage
\onecolumngrid

\appendix

\tableofcontents	

\section{\label{app:site_drive}Analytical properties of the nonreciprocal drive}

Throughout this appendix, we will specialize, for concreteness, to the isotropic Heisenberg chain. The analogous results for other nonreciprocal bipartite spin models following Eq.~\eqref{eq:EOM} are straightforward.

\subsection{\label{app:non_sympl}Proof of the lack of symplectic structure in the dynamics of Eq.~\eqref{eq:EOM}}

Here we demonstrate explicitly that the EOM in Eq.~\eqref{eq:EOM} are not symplectic~\cite{steinigeweg2006symplectic,frank1997geometric}; that is, they cannot be generated by any Hamiltonian. To this end, we first introduce the relevant notions from differential geometry.

The phase space for a system of $L$ classical spins is defined as
\begin{equation}
	\mathcal{P} =\{ \left(\bS_1,\dots,\bS_N\right): |\bS_j|^2=1,\quad \forall j\in\{1,\dots,L\} \},
\end{equation}
with $\bS_j$ a unit vector in three-dimensional space. To incorporate the norm constraint, we can parameterize each spin using its azimuthal angle $\varphi_j$ and its projection on the $z$-axis, $z_j$: 
\begin{equation}
	\bS_j =\left(\sqrt{1-z_j^2}\cos\varphi_j,\; \sqrt{1-z_j^2}\sin\varphi_j,\; z_j\right)^t.
	\label{eq:S-coord}
\end{equation}
Using these coordinates, the symplectic form $\omega$ w.r.t.~the conjugate variables $(\varphi_j,z_j)$, can be locally defined as
\begin{equation}
	\omega = \sum_{j=1}^L \mathrm{d}\varphi_j\wedge \mathrm{d}z_j.
	\label{eq:sympl-form}
\end{equation}
A smooth function $f: \mathcal P\to\mathcal P$ is called \textit{symplectic}, if and only if it preserves the symplectic form, i.e., $f^*\omega=\omega$, where the asterisk denotes the pullback. 
Given an energy function $H$, Hamilton's equations of motion read as
\begin{equation}
	\dot\varphi_j = \partial_{{z}_j} H,\quad \dot z_j =- \partial_{\varphi_j}H.
\end{equation}
A vector field $\boldsymbol{X}$ generates a flow on phase space defined as the solution to: 
\begin{equation}
	\frac{\mathrm d}{\mathrm d t}\bS_j(t) = \boldsymbol{X}(\bS_j(t)).
\end{equation}
Hamilton's equations are associated with the Hamiltonian vector field $\boldsymbol{X}_H$, defined	implicitly by $\iota_{\boldsymbol{X}_H}\omega = \mathrm{d} H$, where $\iota$ is the exterior derivative. The flow generated by $\boldsymbol{X}_H$ is called the Hamiltonian flow. 

A major result in symplectic geometry is that any Hamiltonian flow is symplectic; conversely, if the flow of a complete vector field ${\boldsymbol{X}}$ is symplectic, then $\iota_{\boldsymbol{X}} \omega = \mathrm{d}K$ is a closed form ($\mathrm d\mathrm d K=0$), and hence the flow is locally generated by some Hamiltonian $K$~\cite{steinigeweg2006symplectic}.

\subsubsection{Nonsymplecticity of the exact EOM in Eq.~\eqref{eq:EOM}}

Having introduced these definitions, we can now demonstrate that the flow generated by the EOM~\eqref{eq:EOM} is not symplectic. To do this, without loss of generality we set $J=1$, and consider the vector field $\boldsymbol{X}$ that generates the first half-cycle motion:
\begin{eqnarray}
	\begin{cases}
		\dot\bS_j = \left(\bS_{j-1}+\bS_{j+1}\right)\times\bS_j,\quad & j\ \text{even}\\
		\dot\bS_j =  0,\quad & j\ \text{odd}.
	\end{cases}
	\label{eq:half-cycle}
\end{eqnarray}
We will demonstrate that the form $\iota_X\omega$ is not closed, i.e., $\mathrm d \iota_{\boldsymbol{X}}\omega\neq 0$. Thus, the flow is not locally generated by a Hamiltonian, and hence it is not symplectic. 

To see this, we use the definition in Eq.~\eqref{eq:sympl-form} to calculate
\begin{eqnarray}
	\iota_{\boldsymbol{X}}\omega &=& \sum_{j=1}^L \iota_{\boldsymbol{X}}\left(\mathrm d\varphi_j\wedge\mathrm d z_j\right) 
	= \sum_{j=1}^L \dot\varphi_j\mathrm d z_j - \dot z_j\mathrm d\varphi_j
	= \sum_{j\; \mathrm{even}} \dot\varphi_j\mathrm d z_j - \dot z_j\mathrm d\varphi_j\nonumber\\
	&=& \sum_{j\; \mathrm{even}} \left(z_{j+1} + z_{j-1} 
	- \frac{z_j}{\sqrt{1-z_j^2}}\left[\sqrt{1-z_{j-1}^2}\cos(\varphi_j - \varphi_{j-1}) +\sqrt{1-z_{j+1}^2}\cos(\varphi_j - \varphi_{j+1}) \right] \right)\mathrm d z_j \nonumber\\
	&&\qquad\qquad	- \sqrt{1-z_j^2}\left[\sqrt{1-z_{j-1}^2}\sin(\varphi_j - \varphi_{j-1})+\sqrt{1-z_{j+1}^2}\sin(\varphi_j - \varphi_{j+1}) \right]\mathrm d\varphi_j,
\end{eqnarray}
where in the second line we used Eq.~\eqref{eq:half-cycle} written in the coordinate representation from Eq.~\eqref{eq:S-coord}. A straightforward calculation now gives $\mathrm d\iota_{\boldsymbol{X}}\omega \neq 0$, and hence the flow of ${\boldsymbol{X}}$ is not symplectic. The same argument applies to the second half-cycle. Thus, we conclude that the EOM in Eq.~\eqref{eq:EOM} cannot be generated by a Hamiltonian function.

\subsubsection{Conservation of phase-space volume}

Before, we conclude the discussion, let us also prove that the phase space volume remains conserved under Eq.~\eqref{eq:EOM}. Intuitively, for a fixed half-cycle, each spin is subject to a rotation about a fixed axis, which conserves the phase-space volume; since this is true for both half-cycles, it follows that the entire dynamics preserves the phase space volume, and hence the dynamics are conservative. In the following, we prove this mathematically. 

Formally, the phase space volume is defined by the volume element
\begin{equation}
	\mathrm{d} V = \bigwedge_{j=1}^L \mathrm \omega_j = \mathrm d\varphi_1\wedge\mathrm d z_1 \wedge \cdots\wedge \mathrm d\varphi_L\wedge\mathrm d z_L.
\end{equation}
Liouville's theorem reads as
\begin{equation}
	\label{eq:Liouville}	0 = \mathcal{L}_{\boldsymbol{X}} \mathrm{d} V = \iota_{\boldsymbol{X}}\mathrm d \mathrm{d} V + \mathrm d\iota_{\boldsymbol{X}} \mathrm{d} V =  \mathrm d\iota_{\boldsymbol{X}} \mathrm{d} V,
\end{equation}
where we used that the volume form is closed, $\mathrm d\mathrm{d} V=0$, since the symplectic form itself is closed. Here $ \mathcal{L}_{\boldsymbol{X}}$ denotes the Lie derivative along the flow of ${\boldsymbol{X}}$. 

To demonstrate the validity of Liouville's theorem for Eq.~\eqref{eq:half-cycle}, it suffices to focus on the second spin $\bS_2$:
\begin{eqnarray}
	\mathrm d\iota_{\boldsymbol{X}}\mathrm{d} V &=& \mathrm d \iota_{\boldsymbol{X}}\left(\mathrm d\varphi_1\wedge\mathrm d z_1\wedge\cdots\wedge \mathrm d\varphi_L\wedge\mathrm d z_L \right) \nonumber\\
	&=& \mathrm d\left(z_{3} + z_{1} 
	- \frac{z_2}{\sqrt{1-z_2^2}}\left[\sqrt{1-z_{1}^2}\cos(\varphi_2 - \varphi_{1}) +\sqrt{1-z_{3}^2}\cos(\varphi_2 - \varphi_{3}) \right] \right)
	\mathrm d\varphi_1\wedge\mathrm d z_1\wedge\mathrm d z_2 \wedge\mathrm d\varphi_3\wedge\mathrm d z_3\cdots\wedge \mathrm d z_L \nonumber\\
	&+&	\mathrm d\left( - \sqrt{1-z_2^2}\left[\sqrt{1-z_{1}^2}\sin(\varphi_2 - \varphi_{1})+\sqrt{1-z_{3}^2}\sin(\varphi_2 - \varphi_{3}) \right]\right) 
	\mathrm d\varphi_1\wedge\mathrm d z_1\wedge \mathrm d\varphi_2\wedge\mathrm d\varphi_3\wedge\mathrm d z_3\wedge \cdots\wedge \mathrm d\varphi_L\wedge\mathrm d z_L \nonumber\\
	&& +\; \mathrm{all\;other\;even\; spins} \nonumber\\
	&=& \frac{z_2}{\sqrt{1-z_2^2}}\left[\sqrt{1-z_{1}^2}\sin(\varphi_2 - \varphi_{1}) +\sqrt{1-z_{3}^2}\sin(\varphi_2 - \varphi_{3}) \right] 
	\mathrm d\varphi_2\wedge\mathrm d\varphi_1\wedge\mathrm d z_1\wedge\mathrm d z_2 \wedge\mathrm d\varphi_3\wedge\mathrm d z_3\cdots\wedge \mathrm d\varphi_L\wedge\mathrm d z_L \nonumber\\
	&+&  \frac{z_2}{\sqrt{1-z_2^2}}\left[\sqrt{1-z_{1}^2}\sin(\varphi_2 - \varphi_{1}) +\sqrt{1-z_{3}^2}\sin(\varphi_2 - \varphi_{3}) \right] 
	\mathrm d z_2\wedge\mathrm d\varphi_1\wedge\mathrm d z_1\wedge \mathrm d\varphi_2\wedge\mathrm d\varphi_3\wedge\mathrm d z_3\wedge \cdots\wedge \mathrm d\varphi_L\wedge\mathrm d z_L \nonumber\\
	&& +\; \mathrm{all\;other\;even\; spins} \nonumber\\
	&=&0,
\end{eqnarray} 
where in the last equality we used the antisymmetric property of the wedge product. 

Every symplectic map preserves the phase space volume, but as we have seen, the converse is not true.

\subsection{\label{app:eff_EOM}Details of the auxiliary spin Hamiltonian description}

\subsubsection{Derivation of the EOM for the coupled $S$-$a$ system}

To construct a Hamiltonian for the nonreciprocal drive, we require additional degrees of freedom to impose the nonreciprocity on the original spins; to hold just the odd sites fixed, we require an additional field to cancel the effective field of the even sites -- but without simply measuring the even spins and then applying a site-dependent field (which would obviously not be Hamiltonian). 
We achieve this by coupling the physical spins $\bS$ to a set of auxiliary spins $\ba$. The full system is illustrated in Fig.~\ref{fig:Geometry}b, with the time evolution generated by the Hamiltonian:
\eqn{
	\mH = J\sum_j \bS_j \cdot \bS_{j + 1} + J \sum_j \left(\frac{1}{2} + g(t)(-1)^j\right) \ba_j \cdot (\bS_{j-1}+\bS_{j+1}),
	\label{H_site}
}
where
\eqn{
	g(t) = \frac{1}{2} \mathrm{sgn}(\sin \omega t).
}
It should be noted that we do not fix the dynamics of the $\ba$ spins -- they evolve as unit-length spins under the above Hamiltonian dynamics with their own equations of motion, $\dot\ba = (\pd\mH / \pd \ba) \times \ba$. We will, however, impose a particular set of initial conditions. 

Setting $f_j(t)=\frac{1}{2} + g(t)(-1)^j$, the Hamiltonian equations of motion are:
\eqa{
	\dot\bS_j &= J\left(\bS_{j-1} + \bS_{j+1} + f_{j-1}(t)\ba_{j-1} + f_{j+1}(t)\ba_{j+1}\right) \times \bS_j \nn \\
	\dot\ba_j &= J f_j(t) (\bS_{j-1} + \bS_{j+1}) \times \ba_j.
}
We fix the initial conditions as $\ba_j(0) = -\bS_j(0)$. Now, over the first half-period, $0 < t < \tau/2$, we have $g(t) = +1/2$, which implies
\eqn{
	\dot\ba_j = 
	\begin{cases}
		J (\bS_{j-1} + \bS_{j+1}) \times \ba_j, & j\ \text{even} \\
		0, & j\ \text{odd}
	\end{cases}
}
and
\eqn{
	\dot\bS_j = 
	\begin{cases}
		J\left(\bS_{j-1} + \bS_{j+1}\right) \times \bS_j, & j\ \text{even} \\
		J\left(\bS_{j-1} + \bS_{j+1} + \ba_{j-1} + \ba_{j+1}\right) \times \bS_j, & j\ \text{odd}
	\end{cases}
}

We will now show that, over this half-period, $\dot\bS_j = 0$ for odd $j$ (and so for even $j$, $\bS_j$ evolves in a constant effective field), and, moreover, that the initial condition is preserved, i.e., $\forall j,\; \forall t: \bS_j(t) = -\ba_j(t)$. First, observe that for even $j$, $\bS_j$ and $-\ba_j$ have the same equation of motion with the same initial condition. Thus $\bS_j(t) = -\ba_j(t)$ for even $j$. This then implies that, for odd $j$, the effective field seen by $\bS_j$ vanishes at all times in the half-period, and thus $\dot\bS_j = 0$ for all odd $j$. The equations of motion directly establish $\dot\ba_j = 0$ for odd $j$, and so the initial conditions are preserved. 

This argument is clearly symmetric with respect to the parity of $j$, and so the opposite situation holds over the next half-period. Since the initial condition is preserved throughout, the Hamiltonian (\ref{H_site}) exactly reproduces the nonreciprocal drive.

\subsubsection{\label{app:Magnus}Derivation of the Floquet-Magnus Hamiltonian}

Our goal here is to find the time-independent Floquet Hamiltonian $\mathcal{H}_{F}\left[t_{0}\right]$ which governs the stroboscopic dynamics of the combined system consisting of the original spins $\bS$ and the auxiliary spins $\ba$. Note that $\mathcal{H}_{F}\left[t_{0}\right]$ depends explicitly on the initial choice for the phase of the drive or, equivalently, on the initial time $t_0$, and we keep track of this dependence below. 

Then, using the Floquet-Magnus expansion~\cite{bukov2015universal}, we can construct the Floquet Hamiltonian as a series in the drive-period $\tau$:
\begin{equation}
	\mathcal{H}_{F}\left[t_{0}\right]=\sum_{n=0}^{\infty} \mathcal{H}_{F}^{(n)}\left[t_{0}\right], 
\end{equation}
where the superscript $(n)$ means $\mathcal{H}_{F}^{(n)}\propto\mathcal{O}(\tau^n) = \mathcal{O}(\omega^{-n})$. The lowest-order contributions to the Floquet Hamiltonian are, explicitly,
\begin{eqnarray}
	\mathcal{H}_{F}^{(0)} &=&\mathcal H_{0} \\
	\mathcal{H}_{F}^{(1)}\left[t_{0}\right] &=&\frac{1}{2!\tau}\int_{t_0}^{\tau + t_0} dt_1 \int_{t_0}^{t_1} dt_2 \{H(t_1), H(t_2)\} \\
	\mathcal{H}_{F}^{(2)}\left[t_{0}\right] 
	&=&\frac{1}{3!\tau}\int_{t_0}^{\tau + t_0} dt_1 \int_{t_0}^{t_1} dt_2 \int_{t_0}^{t_2} dt_3 \left( \{H(t_1), \{H(t_2), H(t_3)\}\} + \{H(t_3), \{H(t_2), H(t_1)\}\}\right).
\end{eqnarray}

In the subsequent discussion, it will be convenient to separate out the time-averaged term $\mathcal{H}_0$ and the time-dependent term $\mathcal{V}$,
\begin{eqnarray}
	\mathcal{H}(t) &=& J \sum_{j}\left[\bS_{j} \cdot \bS_{j+1}+ \frac{1}{2} \ba_{j} \cdot\left(\bS_{j-1}+\bS_{j+1}\right)\right] 
	+ g(t) J \sum_{j}(-1)^{j} \ba_{j} \cdot\left(\bS_{j-1}+\bS_{j+1}\right)
	= \mathcal{H}_0 + g(t) \mathcal{V}.
\end{eqnarray}
The explicit time-dependence is now only carried by the function $g(t)$, and the first-order term becomes
\eqn{
	\mathcal{H}_{F}^{(1)}\left[t_{0}\right] =\frac{1}{2!\tau}\int_{t_0}^{\tau + t_0} dt_1 \int_{t_0}^{t_1} dt_2 \left( g(t_1) - g(t_2) \right) \{\mV, \mH_0\} = \left( \frac{t_0}{2} - \frac{\tau}{8} \right)\{\mV, \mH_0\}.
}
A straightforward calculation yields the required Poisson bracket,
\eqn{
	\left\{\mathcal V, \mathcal H_{0}\right\}=J^{2} \sum_{j}(-1)^{j} \varepsilon^{\mu \nu \lambda}\left(\ba_{j}^{\mu}+\ba_{j-2}^{\mu}\right)\left(\bS_{j}^{\nu}+\bS_{j-2}^{\nu}\right) \bS_{j-1}^{\lambda}.
}
Rewriting this in terms of dot and cross products, and shifting some site labels, we arrive at the first-order term,
\eqn{
	\mathcal{H}_{F}^{(1)}\left[t_{0}\right] = \left( \frac{t_0}{2} - \frac{\tau}{8} \right) J^2 \sum_{j}(-1)^{j} \ba_{j} \cdot\left[\left(\bS_{j}+\bS_{j-2}\right) \times \bS_{j-1}+\left(\bS_{j+2}+\bS_{j}\right) \times \bS_{j+1}\right].
}

\subsubsection{\label{app:EOM_eff}Effective EOM}

The equation of motion to order $\omega^{-1}$ can be derived from the effective Hamiltonian $\mathcal H_{F}^{(0)}+\mathcal H_{F}^{(1)}$. We fix the Floquet gauge by setting $t_0 = 0$, and thus obtain
\begin{eqnarray}
	\dot{\ba}_{j} &=& \frac{\partial \mathcal H_{F}^{(0)}}{\partial \ba_{j}} \times \ba_{j}+\frac{\partial \mathcal H_{F}^{(1)}}{\partial \ba_{j}} \times \ba_{j} \nonumber\\ 
	&=&\Big(\frac{J}{2}\left(\bS_{j-1}+\bS_{j+1}\right)-\frac{\tau J^{2}}{8}(-1)^{j}\Big[\left(\bS_{j}+\bS_{j-2}\right) \times \bS_{j-1}
	+\left(\bS_{j+2}+\bS_{j}\right) \times \bS_{j+1}\Big]\Big) \times \ba_{j}.
\end{eqnarray}
Using the conditions $\ba_{j}(t)=-\bS_{j}(t)$, we find
\begin{eqnarray}
	\label{eq.EOM_S}
	\dot{\bS}_{j} =\frac{J}{2}\left(\bS_{j-1}+\bS_{j+1}\right) \times \bS_{j}
	+\frac{\tau J^{2}}{8}(-1)^{j}\Big[\left(\bS_{j}+\bS_{j-2}\right) \times \bS_{j-1} +\left(\bS_{j+2}+\bS_{j}\right) \times \bS_{j+1}\Big] \times \bS_{j}.
\end{eqnarray}

Showing that the initial condition $\ba_j(0) = -\bS_j(0)$ is conserved by the effective dynamics is a simple matter of deriving the general equation of motion for the $\bS$-spins directly, and checking that we obtain the same effective equation of motion (\ref{eq.EOM_S}). We have, to first order,
\begin{eqnarray}
	\dot{\bS}_{j} &=& \frac{\partial\mathcal  H_{F}^{(0)}}{\partial \bS_{j}} \times \bS_{j}+\frac{\partial \mathcal H_{F}^{(1)}}{\partial \bS_{j}} \times \bS_{j} \nonumber\\ 
	&=&\Big(J\left(\bS_{j-1}+\bS_{j+1}\right) + \frac{J}{2}\left(\ba_{j-1}+\ba_{j+1}\right)
	+\frac{\tau J^{2}}{8}(-1)^{j}\Big[\left(\ba_{j}+\ba_{j-2}\right) \times \bS_{j-1}
	\nonumber\\
	&&+\left(\ba_{j+2}+\ba_{j}\right) \times \bS_{j+1} +\left(\ba_{j+1}+\ba_{j-1}\right) \times +\left(\bS_{j+1}+\bS_{j-1}\right) \Big]\Big) \times \bS_{j},
\end{eqnarray}
which, upon inserting the condition $\ba_j = -\bS_j$, may be readily seen to reduce to the same effective equation of motion, Eq.~\eqref{eq.EOM_S}.

In a similar manner, we derive the second-order contribution to the Floquet-Hamiltonian, $\mH_F^{(2)}$. Setting $t_0 = 0$, the effective equations of motion are, to second-order,
\eqa{
	\dot\bS_j &= \frac{J}{2}\left(\bS_{j-1} + \bS_{j+1}\right) \times \bS_j \nn \\
	& - \frac{J^2\tau}{8}(-1)^j \left[ \left( \bS_j + \bS_{j-2} \right) \times \bS_{j-1} + \left( \bS_j + \bS_{j+2} \right) \times \bS_{j+1} \right] \times \bS_j \nn \\
	& + \frac{J^3\tau^2}{96}\biggl( 2\left[\bS_{j-1}\cdot \bS_{j-2}\right]\; \bS_{j-3} \nn \\
	&\qquad\qquad + \left[\bS_{j-1}\cdot(\bS_{j+1} + \bS_j + \bS_{j-2} - 2\bS_{j-3}) - 1\right] \bS_{j-2} \nn \\
	&\qquad\qquad + \left[\bS_j\cdot(\bS_{j+1} + \bS_{j-1} - \bS_{j-2}) + \bS_{j-2}\cdot(\bS_{j-1} - \bS_{j+1}) - 2\right]\bS_{j-1} \nn \\
	&\qquad\qquad + \left[\bS_j\cdot(\bS_{j-1} + \bS_{j+1} - \bS_{j+2}) + \bS_{j+2}\cdot(\bS_{j+1} - \bS_{j-1}) - 2\right]\bS_{j+1} \nn \\
	&\qquad\qquad + \left[\bS_{j+1}\cdot(\bS_{j-1} + \bS_j + \bS_{j+2} - 2\bS_{j+3}) - 1\right] \bS_{j+2} \nn \\ 
	&\qquad\qquad + \left[2\bS_{j+1}\cdot \bS_{j+2}\right]\; \bS_{j+3}\biggr) \times \bS_j \nn\\
	&+ \mO(\tau^3).
}

\subsubsection{\label{app:nonsymplecticEOM_eff}Proof that the effective EOM are nonsymplectic}

The effective equations of motion for the $\bS$-subsystem break symplecticity at first order. To see this, rather than considering the flow of vector fields, let us note that we have terms of the form
\eqn{
\dot{\bS} = (\bS \times \boldsymbol{h}) \times \bS + ...
\label{eq:nonsympterm}
}
in the first-order EOM. For such a term to arise from Hamilton's equations, $\dot{\bS} = \frac{\pd H}{\pd \bS} \times \bS$,
we would require a Hamiltonian $H$ containing a term $K$ such that
\eqn{
\frac{\pd K}{\pd S^{\mu}} = \epsilon^{\mu\nu\lambda} S^{\nu}h^{\lambda}.
}
Up to an irrelevant constant, such a $K$ must be quadratic in $\bS$ and linear in $\boldsymbol{h}$, for which the most generic possibility is
\eqn{
K = A^{\mu\nu\lambda}S^{\mu}S^{\nu}h^{\lambda},
}
for some arbitrary rank-3 tensor $A$. But now we obtain
\eqn{
\frac{\pd K}{\pd S^{\mu}} = (A^{\mu\nu\lambda} + A^{\nu\mu\lambda}) S^{\nu}h^{\lambda},
}
and since $A^{\mu\nu\lambda} + A^{\nu\mu\lambda} = \epsilon^{\mu\nu\lambda}$ is a contradiction [the left-hand-side is symmetric w.r.t.~$\mu\leftrightarrow\nu$, while $\epsilon^{\mu\nu\lambda}$ on the right-hand side is anti-symmetric by definition], we conclude that such terms as (\ref{eq:nonsympterm}) cannot be obtained from a Hamiltonian, and, thus, are nonsymplectic.

\subsection{\label{app:alt_deriv_EOM}Alternative derivations of the effective EOM}

In this appendix, we provide two alternative ways to derive the effective equations of motion using (i) two-times perturbation theory, and (ii) the equation of motion for the phase space density (Liouville's equation). Both methods can be generalized to higher orders in $\tau$ in a straightforward way. 

\subsubsection{\label{app:two_times}Two-times perturbation theory}

\begin{figure}[t!]
	\centering
           \includegraphics[width=\columnwidth]{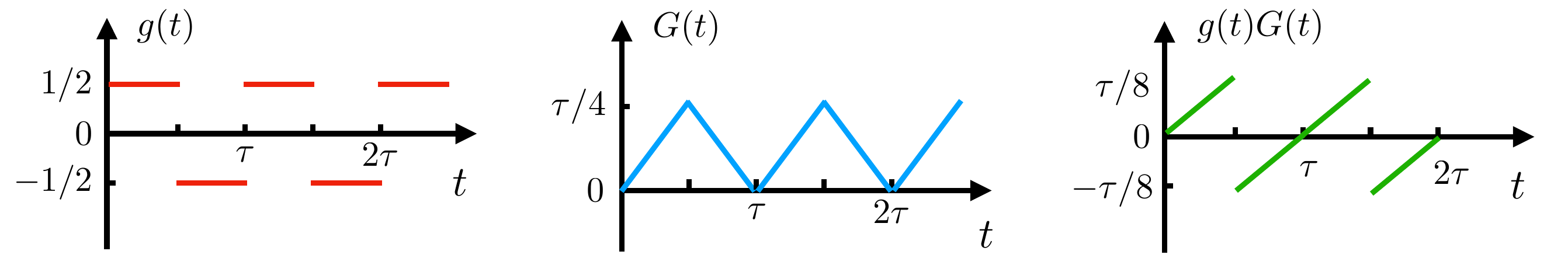}
	\caption{Time dependence of the periodic driving function $g(t)$ (period $
		\tau$) and the anti-derivative $G(t)$, as well as the product $g(t)G(t)$.
	}
	\label{fig:driving_profile}
\end{figure}

Here we illustrate an alternative method via two-times perturbation theory~\cite{strogatz2018nonlinear} to derive the effective EOM for the nonreciprocal drive.

We start by considering the exact EOM, given by:
\begin{eqnarray}
	\label{eq.Floquet_protocol}
	\dot{\bS}_j = -J\left[\frac{1}{2}+(-1)^jg(t)\right]\bS_{j}\times(\bS_{j-1}+\bS_{j+1}), 
\end{eqnarray} 
where $g(t) = \frac{1}{2} \mathrm{sgn}(\sin \omega t)$.
To derive an ansatz for the structure of the fast motion variable, we explicitly separate the time-average from the rest:
\begin{eqnarray}
\label{eq.Floquet_separate}
	\dot{\bS}_j = -\frac{J}{2}\bS_{j}\times(\bS_{j-1}+\bS_{j+1}) + J(-1)^jg(t)\; \bS_{j}\times(\bS_{j-1}+\bS_{j+1}). 
\end{eqnarray}
For fast drivings, the local energy scale separates from the driving frequency, $\omega\gg J$. Therefore, one can decompose the resultant time evolution for spin $\bS_j$ into a fast motion variable (denoted by $\bEta_j$) and slow motion (denoted by $\bsigma_j$). The slow motion reduces to the Heisenberg EOM in the fast driving limit, $\omega\to\infty$, and the fast motion captures additional corrections for any finite $\omega$.

Thus, we consider the following decomposition
\begin{eqnarray}
	\label{eq.decompose}
	\bS_j = \sqrt{1-\bEta_j^2}\bsigma_j+\bEta_j,
\end{eqnarray}
which satisfies the properties $\bEta_j \ll \bsigma_j,\ , |\bsigma_j| = 1,\ \bEta_j \cdot \bsigma_j = 0$. The first relation, $\bEta_j \ll \bsigma_j$, is the statement that the fast-varying field is a small correction to the slow-varying field $\bsigma_j$ in the high-frequency limit; the second condition imposes normalization for the slow variable $\bsigma_j$; together with the first two, the orthogonality relation, $\bEta_j \cdot \bsigma_j = 0$, ensures the normalization of the spin vector $\bS_j$ at all times:
\begin{eqnarray}
    \begin{aligned}
        \bS_j^2= (1-\bEta_j^2)\bsigma_j^2+\bEta_j^2+2\sqrt{1-\bEta_j^2}\bsigma_j\cdot\bEta_j=1.
    \end{aligned}
\end{eqnarray}
Note that a similar decomposition is commonly used in spin wave theory \cite{auerbach1998interacting}. 

The time-average term exhibits slow dynamics by construction, and does not contribute to the fast motion $\bEta_j$; we thus use the remaining term to find the dependence of the fast motion $\bEta_j$ on the slow motion $\bsigma_j$. To do so, we integrate the right-hand side of Eq.~\ref{eq.Floquet_separate} with respect to the fast time, i.e., we treat the slow motion variables $\bsigma_j$ as constant; the leading order contribution to the fast motion is thus given by
\begin{eqnarray}
	\label{eq.eta}
	\bEta_j(t) = -J(-1)^j \left( \int^t_0 \mathrm{d} t' g(t') \right) \bsigma_j\times(\bsigma_{j-1}+\bsigma_{j+1})
	:= -J(-1)^jG(t) \bsigma_j\times(\bsigma_{j-1}+\bsigma_{j+1}).
\end{eqnarray}
Note that $G(t)$, shown in Fig.~\ref{fig:driving_profile}, has the maximum value $\tau/4$, and is thus $\mathcal{O}(\omega^{-1})$ . The fast motion $\bEta_j$ is therefore also $\mathcal{O}(\omega^{-1})$ and, as expected, vanishes as $\omega\to\infty$.

Our goal is to derive an EOM to order $\omega^{-1}$ for the slow motion $\bsigma_j$.
Taking the full time derivative of Eq.~\eqref{eq.decompose} (w.r.t.~the time $t$ which contains both the slow and the fast time variable), one obtains
\begin{eqnarray}
	\label{eq.Eq_sigma}
	\dot{\bsigma}_{j} = \frac{\dot{\bS}_j-\dot{\bEta}_j}{\sqrt{1-\bEta_j^2}}+\frac{\dot{\bEta}_j\cdot \bEta_{j}}{1-\bEta_j^2}\bsigma_j. 
\end{eqnarray}
We now want to eliminate the $\bEta_j$ and ${\bS}_j$ dependence from the right-hand side.
We first use the fact that $\bEta_j$ is $\mathcal{O}(\omega^{-1})$ to simplify the denominators by discarding any terms of higher-order than $\omega^{-1}$ in inverse-frequency, which yields 
\begin{eqnarray}
	\label{eq.eta_order}
	\sqrt{1-\bEta_j^2} = 1 +\mathcal{O}(\omega^{-2}),\qquad {1-\bEta_j^2} = 1 +\mathcal{O}(\omega^{-2}).
\end{eqnarray}
Then we can find an expression for $\dot{\bS}_j$ by inserting the ansatz from Eq.~\eqref{eq.decompose} into Eq.~\eqref{eq.Floquet_protocol}, which leads to 
\begin{eqnarray}
	\begin{aligned}
		\dot{\bS}_j &= -J \left[\frac{1}{2}+(-1)^jg(t)\right]\left( \bsigma_j+\bEta_j\right)\times\left[\left(\bsigma_{j-1}+\bEta_{j-1}\right)+\left(\bsigma_{j+1}+\bEta_{j+1}\right)\right]+\mathcal{O}(\omega^{-2}),
	\end{aligned}
\end{eqnarray}
where we also made use of Eq.~\eqref{eq.eta_order}. 
At the same time, taking the derivative of Eq.~\eqref{eq.eta} w.r.t.~the full time variable (fast and slow), we arrive at
\begin{eqnarray}
	\begin{aligned}
		\dot{\bEta_{j}} &= -J(-1)^jg(t)\;\bsigma_j\times(\bsigma_{j-1}+\bsigma_{j+1})-J(-1)^jG(t)
		\frac{\mathrm d}{\mathrm d t} \left[\bsigma_j\times(\bsigma_{j-1}+\bsigma_{j+1})\right].
	\end{aligned}
\end{eqnarray}
The two equations above for the derivatives $\dot{\bEta_{j}}$ and $\dot{\bS}_j$ can now be inserted in Eq.~\eqref{eq.Eq_sigma}:
\begin{eqnarray}
	\label{eq.Eq_sigma_simple}
	\begin{aligned}
		\dot{\bsigma}_{j} =&\; {\dot{\bS}_j-\dot{\bEta}_j}+{\dot{\bEta}_j\cdot \bEta_{j}}\bsigma_j + \mathcal{O}(\omega^{-2})\\
		=&-J\left[\frac{1}{2}+(-1)^jg(t)\right]\left[\bsigma_j\times(\bsigma_{j-1}+\bsigma_{j+1})+\bEta_j\times(\bsigma_{j-1}+\bsigma_{j+1})+\bsigma_j\times(\bEta_{j-1}+\bEta_{j+1})\right]\\
		&+J(-1)^jg(t)\;\bsigma_j\times(\bsigma_{j-1}+\bsigma_{j+1})+J(-1)^jG(t)\frac{\mathrm d}{\mathrm d t}[\bsigma_j\times(\bsigma_{j-1}+\bsigma_{j+1})]\\
		&+J^2G(t)g(t)\;\bsigma_j[\bsigma_j\times(\bsigma_{j-1}+\bsigma_{j+1})]^2 \\
		&+ \mathcal{O}(\omega^{-2}),
	\end{aligned}
\end{eqnarray}
where, again, Eq.~\eqref{eq.eta_order} has been used. There are two remaining sources of fast motion on the right-hand side of this equation. We eliminate the first, the variable $\bEta_j$, using Eq.~\eqref{eq.eta}. The second source is the time-dependence of the functions $g(t)$ and $G(t)$, which oscillate rapidly in the high-frequency regime $\omega\gg J$ - we thus also need to average over the fast timescale. 

To do this, for any function $h(t)$, we define $\overline{h} = \tau^{-1}\int_0^{\tau}h(t)\mathrm dt$ as the time averaged value over one period $\tau$, and obtain [cf.~Fig.~\ref{fig:driving_profile}]
\begin{eqnarray}
	\overline{g(t)}= 0,\qquad \overline{G(t)} =\frac{\tau}{8},\qquad \overline{g(t)G(t)} = 0,
\end{eqnarray}
see Fig.~\ref{fig:driving_profile}.
Next, we insert Eq.~\eqref{eq.eta} into Eq.~\eqref{eq.Eq_sigma_simple} and perform the time average of the resulting equation over a single period $\tau$. In the limit $\omega\gg J$, one may assume that the slow variable $\bsigma_j$ does not change over the time $\tau$. This yields
\begin{eqnarray}
	\label{eq.eom_sigma}
	\begin{aligned}
		\dot{\bsigma}_{j} =&-\frac{J}{2}\bsigma_j\times(\bsigma_{j-1}+\bsigma_{j+1}) + \frac{(-1)^jJ^2\overline{G}}{2}\;[\bsigma_j\times(\bsigma_{j-1}+\bsigma_{j+1})]\times(\bsigma_{j-1}+\bsigma_{j+1})\\
		&-\frac{J^2(-1)^j\overline{G}}{2}\;\bsigma_j\times [\bsigma_{j-1}\times(\bsigma_{j-2}+\bsigma_j)+\bsigma_{j+1}\times(\bsigma_{j}+\bsigma_{j+2})]\\
		&+J(-1)^j\overline{G}\;[\dot{\bsigma}_{j}\times(\bsigma_{j-1}+\bsigma_{j+1})+\bsigma_{j}\times(\dot{\bsigma}_{j-1}+\dot{\bsigma}_{j+1})] \\
		&+ \mathcal{O}(\omega^{-2}). 
	\end{aligned}
\end{eqnarray}
Finally, in order to derive a self-consistent EOM for ${\bsigma}_{j}$, we need to eliminate $\dot{\bsigma}_{j}$ from the terms on right-hand side. This can be done by noting that all such terms come with an $\mathcal{O}(\omega^{-1})$ prefactor, since $\overline{G}$ is $\mathcal{O}(\omega^{-1})$. 
Hence, the derivatives $\dot\bsigma_j$ on the right-hand-side can be eliminated by using the EOM, Eq.~\eqref{eq.eom_sigma} itself, but in this instance retaining only terms of $\mathcal{O}(1)$,
\begin{eqnarray}
	\begin{aligned}
		\dot{\bsigma}_{j}=-\frac{J}{2}\bsigma_j\times(\bsigma_{j-1}+\bsigma_{j+1})+\mathcal{O}(\omega^{-1}).
	\end{aligned}
\end{eqnarray}
Inserting this back in Eq.~\eqref{eq.eom_sigma}, we arrive at the effective EOM for the slow variable ${\bsigma}_{j}$,
\begin{eqnarray}
	\begin{aligned}
		\dot{\bsigma}_{j} =&\frac{J}{2}(\bsigma_{j-1}+\bsigma_{j+1})\times\bsigma_j-(-1)^j\frac{J^2\tau}{8}[(\bsigma_{j-2}+\bsigma_j)\times\bsigma_{j-1}+(\bsigma_j+\bsigma_{j+2})\times\bsigma_{j+1}]\times\bsigma_j+\mathcal{O}(\omega^{-2}).
	\end{aligned}
\end{eqnarray}

\subsubsection{\label{app:Liouville_eff}Effective Liouville equation approach}

The modern theoretical analysis of prethermalization is based on Floquet's theorem, which requires the linearity of the equations of motion (EOM)~\cite{kuchment1993floquet}. Since thermalization can microscopically be traced back to chaotic trajectories, and chaos in classical systems can only occur in nonlinear EOM~\cite{d2016quantum}, this might at first appear paradoxical. A similar ``problem'' with Floquet's theorem occurs for quantum dynamics in the Heisenberg picture, where the EOM are also nonlinear. The application of Floquet's theorem in these cases is justified by the linearity of the alternative Schr\"odinger picture based on the Liouville-von Neumann equation for the density operator (phase-space density)~\cite{higashikawa2018floquet}, which is defined by means of a commutator (Poisson bracket) structure in Hamiltonian mechanics~\cite{mori2018floquet}. For classical systems, it is equivalent to the existence of conjugate variables, and induces a symplectic structure on phase space.

Yet a third way to obtain the effective EOM in a systematic expansion controlled by $\omega^{-1}$ is as follows: note that the Liouville equation, Eq.~\eqref{eq:Liouville}, is a linear, time-periodic ODE for the phase space density. Although, for our system, it cannot be written in its familiar form using the Poisson bracket due to the lack of a symplectic structure, the latter is not required to apply Floquet's theorem.

Indeed, by exploiting this fact, it was shown in Ref.~\cite{higashikawa2018floquet} that the Floquet-Magnus expansion can be applied to differential equations of the form
\begin{equation}
	\dot{\vec{\bS}} = \vec{\boldsymbol{X}}(\vec{\bS},t),   \qquad \vec{\boldsymbol{X}}(\vec{\bS},t+\tau)=\vec{\boldsymbol{X}}(\vec{\bS},t),
\end{equation} 
where $\vec{\bS}$ denotes all spin variables of the system, and $\vec{\boldsymbol{X}}$ is a (possibly nonlinear) but time-periodic vector field flow. 

The slow dynamics of the system is then captured by the effective EOM
\begin{equation}
	\dot{\vec{\bS}} = \vec{\boldsymbol{X}}_\mathrm{eff}(\vec{\bS});
\end{equation}
To leading order in the inverse-frequency, the Floquet-Magnus expansion, we have
\begin{eqnarray}
	\vec{\boldsymbol{X}}_\mathrm{eff}&=&\sum_{n=0}^\infty \vec{\boldsymbol{X}}_\mathrm{eff}^{(n)},\qquad\qquad \vec{\boldsymbol{X}}_\mathrm{eff}^{(n)} \propto \omega^{-n}, \nonumber\\
	\vec{\boldsymbol{X}}_\mathrm{eff}^{(0)} &=& \frac{1}{\tau}\int^t_0\mathrm d t \vec{\boldsymbol{X}}(\vec{\bS},t),  \nonumber\\
	\vec{\boldsymbol{X}}_\mathrm{eff}^{(1)}  &=&  \frac{1}{4\pi\omega}\int^t_0\mathrm d t_1\int_0^{t_1}\mathrm d t_2 \left[ \vec{\boldsymbol{X}}(\vec{\bS},t_1),\vec{\boldsymbol{X}}(\vec{\bS}, t_2) \right]_\mathcal{L}  ,
\end{eqnarray} 
where the Lie bracket $\left[\cdot, \cdot\right]_\mathcal{L}$ of two vector fields $\vec{\boldsymbol{X}}(\vec{\bS})$ and $\vec{\boldsymbol{Y}}(\vec{\bS})$ is defined as
\begin{equation}
	\mathcal{L}_{\vec{\boldsymbol{X}}}\vec{\boldsymbol{Y}}
	= \left[\vec{\boldsymbol{X}}(\vec{\bS}), \vec{\boldsymbol{Y}}(\vec{\bS}) \right]_\mathcal{L} = 
	\vec{\boldsymbol{X}}(\vec{\bS})\cdot \vec\nabla_{\vec{\bS}} \vec{\boldsymbol{Y}}(\vec{\bS})
	- 	\vec{\boldsymbol{Y}}(\vec{\bS})\cdot \vec\nabla_{\vec{\bS}} \vec{\boldsymbol{X}}(\vec{\bS}).
\end{equation}
Performing the derivatives and calculating the time-ordered integral, we arrive at the effective EOM in Eq.~\eqref{eq:EOM-eff}.

\section{\label{app:bond_drive}Comparison between Hamiltonian (reciprocal) and nonreciprocal drives}

In the main text we have analyzed a nonreciprocal drive for the classical Heisenberg chain, given by the successive evolution of even and odd-numbered sites. A natural complement of this is to consider a drive given by the successive evolution of even and odd-numbered \textit{bonds}.

\begin{figure}
    \centering
    \includegraphics{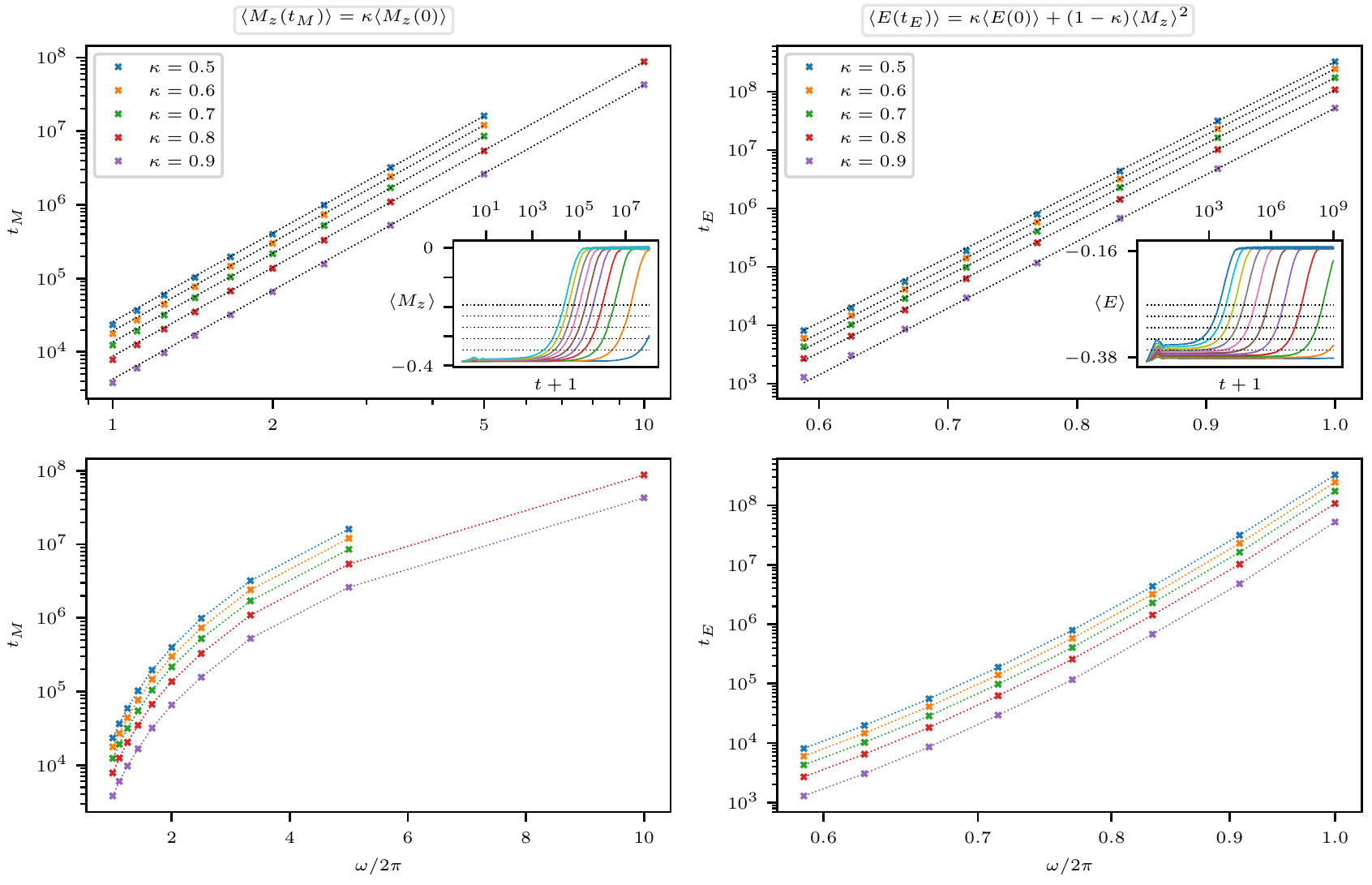}    
\caption{Comparison between the magnetization loss in the \textit{nonreciprocal} drive (left) and Floquet heating in the \textit{reciprocal} (Hamiltonian) drive (right), for the same initial ensemble ($2000$ states, $\beta = 1$, $L = 128$). (Upper panels) Both drives exhibit a prethermal plateau (see insets), but heating in the reciprocal drive is suppressed exponentially, compared to the power-law suppression in the nonreciprocal drive. Inset for the nonreciprocal drive shows the curves from $\tau = 0.1$ (rightmost) to $\tau = 1.0$ (leftmost); inset for the reciprocal drive shows the curves from $\tau = 0.8$ (rightmost) to $\tau = 1.8$ (leftmost). The functional form of the suppression is independent of the threshold value (controlled by $\kappa$) used to define $t_M$ and $t_E$ -- the dotted lines in the insets correspond to $\kappa = 0.5$ (top) to $\kappa = 0.9$ (bottom). (Lower panels) show convincingly that the nonreciprocal drive is not described by an exponential suppression, and, conversely, that the reciprocal drive is not described by a power-law suppression.}
    \label{fig:sym-non-sym-comp}
\end{figure}

In contrast to the nonreciprocal site-based drive considered in the main text, the bond-based drive \textit{is} reciprocal, and is generated by the time-dependent Hamiltonian
\eqn{
	H_\mathrm{rec}(t) = J \sum_j \left( \frac{1}{2} + (-1)^j g(t) \right) \bS_j \cdot \bS_{j+1}, 
	\label{eq:H_bond}
}
which has the same infinite-frequency limit $H_\mathrm{\infty}$ as the nonreciprocal drive; the two drives differ in the way they approach the infinite-frequency limit.

As in the nonreciprocal case, each step of the reciprocal drive is exactly solvable. In contrast to the nonreciprocal drive, the reciprocal drive exactly conserves the magnetization; it does not, however, conserve the energy $H_\mathrm{\infty}$ (at finite frequency). This is the canonical situation to which rigorous estimates for the rate of energy absorption~\cite{mori2018floquet} apply, and we numerically verify these predictions.

We take the same initial ensemble at $\beta = 1$ as we used for the nonreciprocal evolution in the main text. In Fig.~\ref{fig:sym-non-sym-comp}, we find that the system heats up to the maximum entropy state consistent with the conserved magnetization (i.e., $\avg{E} \rightarrow J\avg{M}^2$), and that, as expected from Ref.~\cite{mori2018floquet}, the heating time $t_E$ is exponentially suppressed with the driving frequency, $t_E \sim \exp(-c\,\omega)$, for some model-dependent constant $c$. This stands in stark contrast to the algebraic suppression observed for the nonreciprocal drive in the main text.


\section{\label{app:other_systems}Time-reversal-symmetry breaking drives -- the triangular lattice}

The bipartite nonreciprocal drive from Eq.~\eqref{eq:EOM} considered in the main text and above preserve time-reversal symmetry, in the sense that there exists a Floquet gauge $t_0$~\cite{bukov2015universal} in which the drive is symmetric under time-reversal. This implies that the first-order correction in the van Vleck IFE must vanish~\cite{bukov2015universal}, and hence $\mH_{\mathrm{eff}} \sim \mH_{\mathrm{eff}}^{(0)} + \mH_{\mathrm{eff}}^{(2)}$. 
As discussed in the main text, this may lead one to the incorrect conclusion that the $\omega^4$-scaling arises perturbatively in the IFE using Fermi's Golden rule. 

As we explain in detail below, however, this is not the case.
To see why, notice first that according to Floquet theory, drives that break time-reversal symmetry necessarily have a non-vanishing first-order correction $\mH_{\mathrm{eff}}^{(1)}$~\cite{bukov2015universal}. Fermi's Golden rule would then naively imply an $\omega^2$ scaling. 
However, the triangular lattice Heisenberg model, under a time-reversal-symmetry breaking drive, that we discuss below in detail, exhibits an $\omega^4$ prethermal behavior (Fig.~\ref{fig:sdcq_M}, triangles).
This shows that the $\omega^4$ law is not captured by the perturbative IFE. 

\begin{figure}
    \centering
    \includegraphics[width=\textwidth]{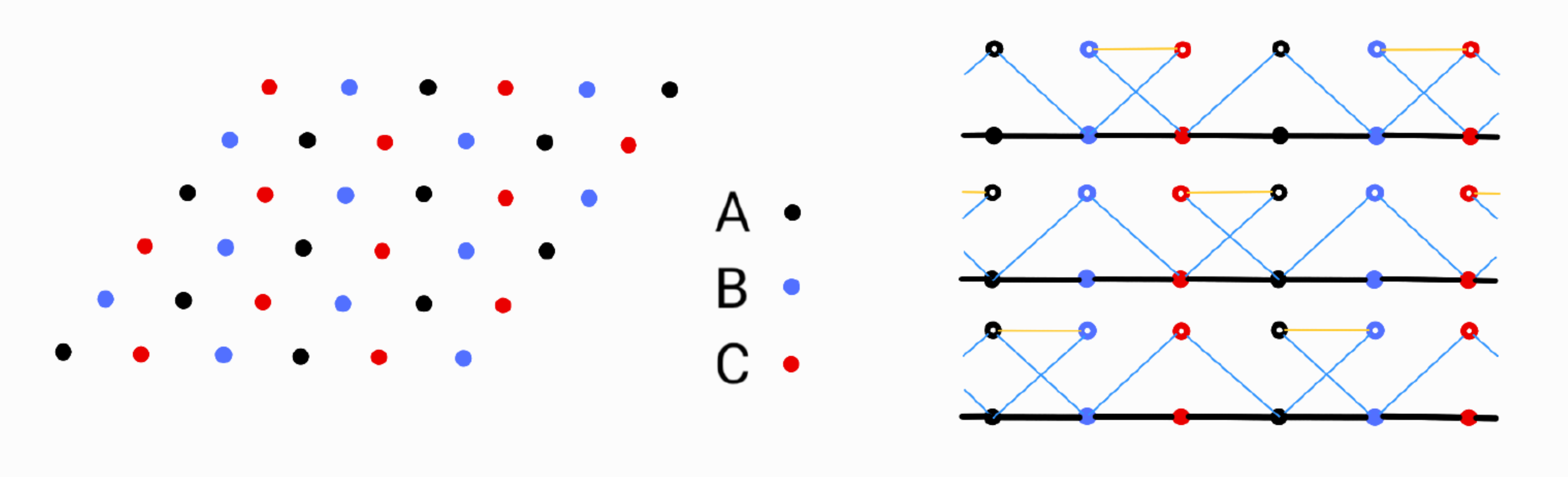}
    \includegraphics{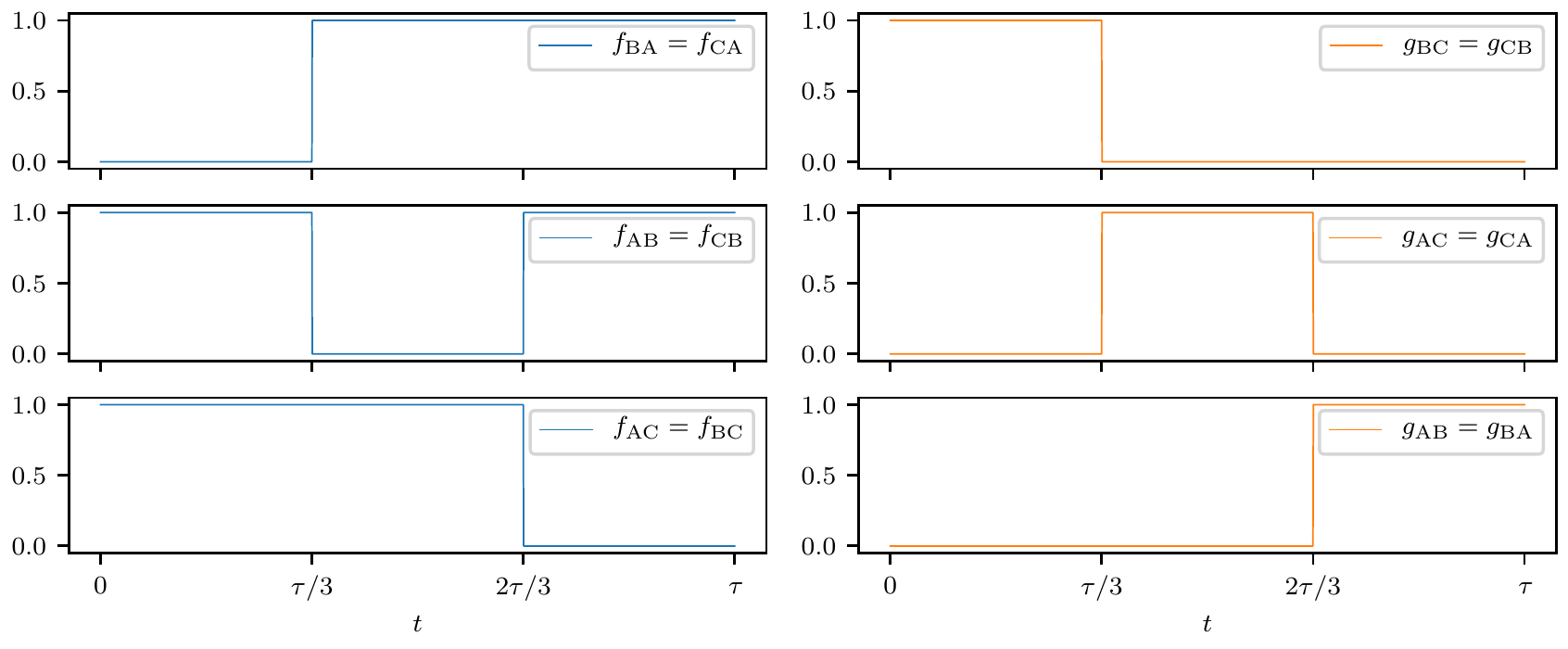}
    \caption{The tripartite drive for the triangular lattice. (Top left): an example tripartite decomposition of the triangular lattice. (Top right): configuration of couplings $f$ and $g$ (blue and orange lines, respectively), for each step of the drive ($\mathcal{A}\mathcal{B}\mathcal{C}$, from the top down); the black lines show the $\bS-\bS$ couplings, which are always on. $\bS$-degrees of freedom are represented by filled circles; the auxiliaries by empty circles. (Lower panels): the values of the couplings $f$ and $g$ across one drive-period for the tripartite drive.
    }
    \label{fig:SI_couplings}
\end{figure}

The general tripartite nonreciprocal periodic drive has the equations of motion:
\begin{eqnarray}
	\label{eq:tripartite_EOM}
	\begin{array}{lr}
		\begin{cases}
			\dot\bS_j^{\mu} = \epsilon^{\mu\nu\lambda} \left(\sum_i J_{ij}^{\nu}\bS_{i}^{\nu} \right)\bS_j^{\lambda},\quad & j \in \mathcal{A}\\
			\dot\bS_j =  0,\quad & j \in \mathcal{B}\\
            \dot\bS_j =  0,\quad & j \in \mathcal{C}
		\end{cases};\quad
		&\text{for}\; t\in\left[0,\frac{\tau}{3}\right), \\ \\
        \begin{cases}
			\dot\bS_j =  0,\quad & j \in \mathcal{A}\\
            \dot\bS_j^{\mu} = \epsilon^{\mu\nu\lambda} \left(\sum_i J_{ij}^{\nu}\bS_{i}^{\nu} \right)\bS_j^{\lambda},\quad & j \in \mathcal{B}\\
			\dot\bS_j =  0,\quad & j \in \mathcal{C}
		\end{cases};\quad
		&\text{for}\; t\in\left[\frac{\tau}{3},\frac{2\tau}{3}\right), \\ \\
		\begin{cases}
			\dot\bS_j = 0,\quad & j \in \mathcal{A}\\
            \dot\bS_j =  0,\quad & j \in \mathcal{B}\\
			\dot\bS_j^{\mu} = \epsilon^{\mu\nu\lambda} \left(\sum_i J_{ij}^{\nu}\bS_{i}^{\nu} \right)\bS_j^{\lambda},\quad & j \in \mathcal{C}
		\end{cases};\quad
		&\text{for}\; t\in\left[\frac{2\tau}{3},\tau\right), 
	\end{array}
\end{eqnarray}
where $\mathcal{A}$, $\mathcal{B}$, $\mathcal{C}$ denote the three sublattices, cf.~Fig.~\ref{fig:SI_couplings}. The $\mathcal{ABCABC}$ pattern 
of this drive (where each step is labeled by the sublattice which is evolving) violates time-reversal because there is no time $t_0$ about which this drive is even -- time-reversal will always flip this to a $\mathcal{CBACBA}$ pattern. By contrast, the bipartite drive, as written in Eq.~(\ref{eq:EOM}), \textit{is} time-reversal symmetric about $t_0 = \tau/4$.

Like Eq.~\eqref{eq:EOM}, this nonreciprocal drive is not generated from any Hamiltonian for the $\bS$-degrees of freedom alone. Again, however, we may use the auxiliary degrees of freedom to construct a Hamiltonian amenable to analysis by IFE. This is more involved than in the bipartite case, since we now require couplings between different auxiliaries. We write the Hamiltonian of the extended system as
\eqn{
\mH(t) = \sum_{i,j} J_{ij} \bS_i \cdot \bS_j + \sum_{i,j} \ba_i \cdot \left( f_{ij}(t)\bS_j + g_{ij}(t)\ba_j \right).
}
Note that $f_{ij}(t) \neq f_{ji}(t)$. We show the values of these couplings in Fig.~\ref{fig:SI_couplings}.

Now, following Ref.~\cite{sun2022engineering}, the first-order term in the van Vleck IFE is
\eqn{
\mH_{\mathrm{eff}}^{(1)} = \frac{1}{2!\tau}\int_0^{\tau}dt_1 \int_0^{t_1}
dt_2 \;\;\frac{2}{\tau} \left[\left(\frac{\tau}{2} - (t_1 - t_2)\right)\mathrm{mod}\tau \right] \{\mH(t_1), \mH(t_2)\}.
}
A straightforward calculation shows that the relevant Poisson bracket can be calculated as
\eqa{
\{\mH(t_1), \mH(t_2)\} = \sum_{ijl} \epsilon^{\mu\nu\lambda}\biggl(& J_{ij} \bigl[ f_{lj}(t_2) - f_{li}(t_2) - f_{lj}(t_1) + f_{li}(t_1) \bigr] a_l^{\nu}S_i^{\mu}S_j^{\lambda} \nn \\
&+ f_{lj}(t_1)f_{il}(t_2) a_i^{\mu} S_j^{\nu}S_l^{\lambda} \nn \\
&+ \bigl[ f_{il}(t_1)f_{jl}(t_2) - 2f_{il}(t_1)g_{ij}(t_2) - 2g_{ij}(t_1)f_{il}(t_2) \bigr] a_i^{\mu} a_j^{\nu}S_l^{\lambda} \nn \\
&+\bigl[ 2g_{ij}(t_1)g_{il}(t_2) - 2g_{ij}(t_1)g_{jl}(t_2)\bigr] a_i^{\mu} a_j^{\nu}a_l^{\lambda} \biggr).
}

To show that $\mH_{\mathrm{eff}}^{(1)}\neq 0$, it suffices to show that any one of the unlike terms is nonzero, so we focus on the $a$-$a$-$a$ term,
\eqn{
\sum_{ijl} \epsilon^{\mu\nu\lambda} \bigl[ 2g_{ij}(t_1)g_{il}(t_2) - 2g_{ij}(t_1)g_{jl}(t_2)\bigr] a_i^{\mu} a_j^{\nu}a_l^{\lambda}.
}
Now, under the sum, we must have that $i, j, l$ are distinct sites, or the antisymmetric tensor will kill the term. Without loss of generality, let $i \in \mathcal{A}$. Since $\mA$ sites are not nearest-neighbours of $\mA$ sites, we must then have $j \in \mathcal{B}, \mathcal{C}$ (or $g_{ij} = 0$ will kill the term); again, without loss of generality, let $j \in \mathcal{B}$.

Now, \textit{a priori}, $l$ may belong to any sublattice, so long as $l \neq i, j$. The various cases yield:
\eqn{
\begin{cases}
    l \in \mathcal{A}: & 2g_{AB}(t_1)g_{AA}(t_2) - 2g_{AB}(t_1)g_{BA}(t_2) = -2g_{AB}(t_1)g_{BA}(t_2) \\
    l \in \mathcal{B}: & 2g_{AB}(t_1)g_{AB}(t_2) - 2g_{AB}(t_1)g_{BB}(t_2) = 2g_{AB}(t_1)g_{BA}(t_2) \\
    l \in \mathcal{C}: & 2g_{AB}(t_1)g_{AC}(t_2) - 2g_{AB}(t_1)g_{BC}(t_2) = 0. \\
\end{cases}
}
The relevant time-ordered integral in the IFE, therefore, is
\eqn{
\frac{1}{2!\tau}\int_0^{\tau}dt_1 \int_0^{t_1}
dt_2 \;\;\frac{2}{\tau}\left[\left(\frac{\tau}{2} - (t_1 - t_2)\right)\mathrm{mod}\tau \right] g_{AB}(t_1) g_{AB}(t_2) = \frac{7\tau}{648} \neq 0.
}
This implies, at least, that the $a$-$a$-$a$ terms do not vanish. 

It follows that $\mH_{\mathrm{eff}}^{(1)} \neq 0$, as expected in the absence of time-reversal symmetry, and that the $\omega^4$-scaling of the prethermal lifetime in the nonreciprocal periodically-driven spin dynamics cannot be explained perturbatively using the IFE.

\section{\label{app:numerics}Details of the numerical simulations}

In this appendix we provide further details of our numerical procedures. We first discuss the construction of the initial states in the thermal ensembles, and then give an overview of the integration of the equations of motion.

\subsection{Initial ensemble}

The initial states are constructed using heatbath Monte Carlo (MC) simulations \cite{loison2004canonical}, which uses the fact that the thermal distribution of a single spin $\bS_i$ in its local field $\bS_{i-1} + \bS_{i+1} + h\hat{\boldsymbol{z}}$ is exactly invertible (see also supplementary material of Ref.~\cite{mcroberts2022anomalous}). We use ensembles of $2000$ states, and each state begins as a completely independent random configuration. We then perform $L \times 10^5$ heatbath updates (randomly selecting the spin to be redrawn from its local thermal distribution) to cool the state to the desired temperature. We use the same set of $2000$ initial states for all dynamical evolution protocols (for a fixed $L$ and $\beta$), to ensure a fair comparison between the reciprocal and nonreciprocal drives, and between the exact nonreciprocal dynamics and the dynamics given by the effective Floquet-Magnus Hamiltonian; we have checked that using different initial ensembles does not change the reported results.

\subsection{Dynamical evolution}

We now turn to the details of the dynamical evolution. The nonreciprocal drive, Eq.~\eqref{eq:EOM}, and the reciprocal drive in Eq.~\eqref{eq:H_bond}, can be integrated to machine precision, since the exact solution can be written in closed form for each step of the drive: a single spin in a constant magnetic field evolves as
\eqn{
	\dot{\bS} = \boldsymbol{M} \times \bS \;\; \Rightarrow\;\; \bS(t) = \exp(\boldsymbol{M}\cdot\boldsymbol{R} \;\;t) \bS(0),
}
where $\boldsymbol{R}$ denotes the (vector of) the generators of rotations. For the nonreciprocal drive, half of the spins $\bS_i$ evolve in the constant (over the half-period) field $\bS_{i-1} + \bS_{i+1}$; for the reciprocal drive, the spins on a bond $\{i, i+1\}$ evolve in the constant field $\bS_i + \bS_{i+1}$.

The fact that each step of the drive is exactly solvable means that the numerical evolution is very efficient 
and accumulates only machine precision errors, allowing us to evolve to long times $t_f = 10^8$, or even $t_f = 10^9$. The values of the energy $H_{\infty}$ and the magnetization $\bM_z$ are stored at $10^4$ stroboscopic times on a log-spaced (to the nearest stroboscopic time $t \in \tau\mathbb{Z}$) grid.

In contrast, the dynamics generated by the effective Hamiltonians $\mH_F^{(1)}$ and $\mH_F^{(2)}$ are not exactly integrable, and we use the standard fourth-order Runge-Kutta (RK4) method with a timestep of $\delta t = 0.001$ (in units of $|J|{=}1$). We store the values of the observables at the stroboscopic times on the log-spaced grid used for the exact dynamics, up to the final time of the RK4 simulations, $t_f = 10^6$. With these values of $\delta t$ and $t_f$, the typical error in the energy density over the simulations (which should be conserved by the nonreciprocal drive and its effective Hamiltonians) is $\sim 10^{-12}$.

\section{\label{app:quantization}Contradictions in the canonical quantization of the nonreciprocal periodic drive}


In this section, we argue that a physical quantum version of the nonreciprocal periodic drive in Eq.~\eqref{eq:EOM} does \textit{not} exist. In particular, we show that there exists no completely positive trace-preserving (CPTP) time-periodic map, such that: 
(i) in the infinite-frequency limit, the dynamics reduces to that of the quantum Heisenberg model, and
(ii) in the classical limit, the dynamics reduces to Eq.~\eqref{eq:EOM}. 

It suffices to restrict to  a two-spin system and set $J=1$. Now, the classical Liouville equation for the phase space density $\rho(\bS_1,\bS_2;t)$, which evolves following the flow field corresponding to Eq.~\eqref{eq:EOM}, is
\begin{eqnarray}
    \label{eq:EOM_Liouville}
    \begin{cases}
			\partial_t\rho(\bS_1,\bS_2;t) = S_2^\alpha \{S_1^\alpha, \rho(\bS_1,\bS_2;t) \},\qquad \; t\in\left[0,\frac{\tau}{2}\right),  \\
			\partial_t\rho(\bS_1,\bS_2;t) = S_1^\alpha \{S_2^\alpha, \rho(\bS_1,\bS_2;t) \},\qquad \; t\in\left[\frac{\tau}{2},\tau\right),
   \end{cases}
\end{eqnarray}
where, $\{\cdot,\cdot\}$ denotes the Poisson bracket.
These equations satisfy condition (i) above, which can be seen by taking the time-average. 

Naively, quantizing Eqs.~\eqref{eq:EOM_Liouville} is straightforward: one replaces the spin variables by the corresponding operators, $\{\cdot,\cdot\}\mapsto -i [\cdot,\cdot]$, and the product of two functions by half their anticommutator, $f(\bS_1,\bS_2)g(\bS_1,\bS_2)\mapsto \frac{1}{2}[f(\bS_1,\bS_2),g(\bS_1,\bS_2)]_+$. This leads to a von Neumann-like equation for the quantum density matrix:
\begin{eqnarray}
    \label{eq:EOM_vonNeumann}
    \begin{cases}
			\partial_t\hat{\rho}(t) = -\frac{i}{2}[S_2^\alpha, [S_1^\alpha, \hat{\rho}(t) ]]_+\;,\qquad &\text{for}\; t\in\left[0,\frac{\tau}{2}\right),  \\
			\partial_t\hat{\rho}(t) = -\frac{i}{2}[S_1^\alpha, [S_2^\alpha, \hat{\rho}(t) ]]_+\;,\qquad &\text{for}\; t\in\left[\frac{\tau}{2},\tau\right).
   \end{cases}
\end{eqnarray}
Note that the second equation is equivalent to the first under the exchange of the spin variables, $\bS_1\leftrightarrow\bS_2$, as expected.

Next, we rewrite these equations to single-out the period-averaged contribution:
\begin{eqnarray}
    \label{eq:EOM_vonNeumann}
    \begin{cases}
			\partial_t\hat{\rho}(t) = -\frac{i}{2}[ S_1^\alpha S_2^\alpha, \hat{\rho}(t) ]
            -\frac{i}{2}\left( S_2^\alpha \hat{\rho} S_1^\alpha - S_1^\alpha \hat{\rho} S_2^\alpha  \right)
   \;,\qquad &\text{for}\; t\in\left[0,\frac{\tau}{2}\right),  \\
			\partial_t\hat{\rho}(t) = -\frac{i}{2}[ S_1^\alpha S_2^\alpha, \hat{\rho}(t) ]
            -\frac{i}{2}\left( S_1^\alpha \hat{\rho} S_2^\alpha - S_2^\alpha \hat{\rho} S_1^\alpha  \right) \;,\qquad &\text{for}\; t\in\left[\frac{\tau}{2},\tau\right).
   \end{cases}
\end{eqnarray}
Clearly, Eqs.~\eqref{eq:EOM_vonNeumann} obey conditions (i) and (ii). Moreover, it is straightforward to check that Eqs.~\eqref{eq:EOM_vonNeumann} preserve the trace of the density matrix $\hat{\rho}$. 
However, Eqs.~\eqref{eq:EOM_vonNeumann} do not define completely positive maps -- that is, some eigenvalues of $\hat{\rho}(t)$ may become negative; a contradiction if the density matrix is to be interpreted as a probability distribution. To see this, it suffices to focus on the first half-cycle, and re-write the second term as 
\begin{equation}
    -\frac{i}{2}\left( S_2^\alpha \hat{\rho} S_1^\alpha - S_1^\alpha \hat{\rho} S_2^\alpha  \right)
     = 
     -\frac{i}{2}\left( S_2^\alpha \hat{\rho} S_1^\alpha - S_1^\alpha \hat{\rho} S_2^\alpha - \frac{1}{2}[S_1^\alpha S_2^\alpha,\hat{\rho}]_+ + \frac{1}{2}[S_1^\alpha S_2^\alpha,\hat{\rho}]_+  \right)
     =: \sum_{m,n} h_{mn}\left( L_m\hat{\rho} L^\dagger_n - \frac{1}{2}[L_n^\dagger L_m,\hat{\rho}]_+ \right),
\end{equation}
with $L_m = (S_1^x,S_1^y,S_1^z, S_2^x,S_2^y,S_2^z)_m$ and $L_n^\dagger = (S_2^x,S_2^y,S_2^z, S_1^x,S_1^y,S_1^z)_n$.
One can convince oneself that, with the above definition, $h_{mn} = h_{nm}^* \in i\mathbb{R}$ is both Hermitian and purely imaginary. Hence, the eigenvalues of $h$ come in pairs, $(-\lambda_i, \lambda_i)$, of positive and negative numbers. This implies that $h$ is not positive semidefinite, and thus the above equation does not define a completely positive map, despite its formal similarity with the Lindblad master equation.
It follows that the quantized equations of motion, Eq.~\eqref{eq:EOM_vonNeumann}, are not CPTP maps, and thus do not govern the dynamics of a physical quantum system. 

This implies the nonreciprocal drive provides an example of a system with a well-defined classical phase space dynamics (via Eq.~\eqref{eq:EOM_Liouville}) for which no quantum equivalent exists. Recently, it was demonstrated that periodically-driven open quantum systems do not always possess a Floquet Lindbladian~\cite{schnell2020is}, which might be related to symplecticity breaking.

\end{document}